# Service Embedding in IoT Networks

Haider Qays Al-Shammari, Ahmed Lawey, Taisir El-Gorashi and Jaafar M. H. Elmirghani
University of Leeds, UK

***Abstract*— The Internet of Things (IoT) is anticipated to participate in the execution of a variety of complex tasks in the near future. IoT objects capable of handling multiple sensing and actuating functions are the cornerstone of smart applications such as smart buildings, smart factories, home automation, and healthcare automation. These smart applications express their demands in terms of high-level requests. These requests are characterised by the different requirements of sensing/actuating functions, processing and memory needs, activation zones, latency, etc. In service-oriented architecture-based IoT, application requests are translated into a business process (BP) workflow. In this study, we model such a BP as a virtual network containing a set of virtual nodes and links connected in a specific topology. These virtual nodes represent the requested processing and location where sensing or actuation are needed. The virtual links capture the requested communication requirements between nodes. In this paper, we introduce a framework, optimised using mixed integer linear programming (MILP), that embeds the BPs from the virtual layer into a lower-level implementation at the IoT physical layer. The proposed framework results in a physical plan that optimally allocates the processing needed and provisions the sensing and actuation at the required locations requirements to an appropriate set of IoT nodes. The optimisation goal is to minimise the IoT layer's total power consumption and optimise the traffic distribution in a manner that minimises the traffic latency of each IoT node. Our results show that the proposed framework enhances the network performance by reducing the power consumption and latency.**

## INTRODUCTION

In the near future, a considerably large number of physical objects will contain sensors and actuators and will have the ability to communicate, forming the basis for the Internet of Things (IoT) [1]. IoT has motivated many global establishments to research and invest in this area and in its promising use in healthcare, transportation, and other smart building applications [2]. However, these promises of IoT come with considerable challenges. One of these challenges is the energy used and its effects on the environment and the expenditure involved [3], [4]. Although each IoT device consumes low power, it is predicted that the number of IoT nodes will reach approximately 50 billion by the year 2020 [5], a massive number that can cause a high aggregate power consumption, as smart cities and smart building applications for example are expected to use a large number of IoT devices across cities [6]. Therefore, minimising the energy consumed by such applications can play a significant role in reducing the total energy consumed by IoT.

In this study, we investigated solutions that can enable IoT to enhance real-world applications in a smart building. A smart building setup consists of a system for the monitoring and control of certain specific applications in the public or private areas of the building. The monitoring and control system consists of distinct types of sensors and actuators such as motion detectors, sound detectors, light detectors, smoke detectors, alarms, and gate controllers. These sensors and actuators, by means of wireless nodes, are connected in the building through a mesh topology. In a smart building, there are distinct applications that use the same resources in the monitoring and control system. For example, both a security application and an energy saving application may use motion detectors, radio frequency identification (RFID) tags, and light detectors for monitoring simultaneously. To be successful, the smart building concept needs to be supported by various applications and has to be adopted by a range of industries, service providers, and administrations. It has to be applied efficiently in a mutual pattern for these sectors [7]. An essential phase toward the realisation of the smart building concept involves the improvement of the communication infrastructure. The IoT paradigm is capable of collecting data from a massive assortment of distinct devices uniformly and seamlessly. The decentralised and heterogeneous properties of IoT devices capable of providing multiple functions require an efficient architecture that hides such heterogeneity from higher-level applications and provides interoperability for information exchange with other IoT devices [8]. A Service Oriented Architecture (SOA) is potentially a viable middleware between users' applications and the IoT physical layer and can achieve interoperability between these heterogeneous IoT devices [9]. SOA enables the abstraction of IoT device functions that can then be translated into basic services, which in turn can be composed of complex services and exploited in the upper application layer. Figure 1 depicts the SOA middleware for IoT, which is composed of three sub-layers [1], [2], [10]: (i) object abstraction layer that enables IoT devices to provide their functions to the upper layers; (ii) service management layer to enable dynamic object discovery, status monitoring, and mapping of the available services to the IoT devices' abstracted functions; and (iii) service composition layer where complex services, referred to as the business process (BP) workflow, are created from the basic services provided by the service management layer [11, 12].

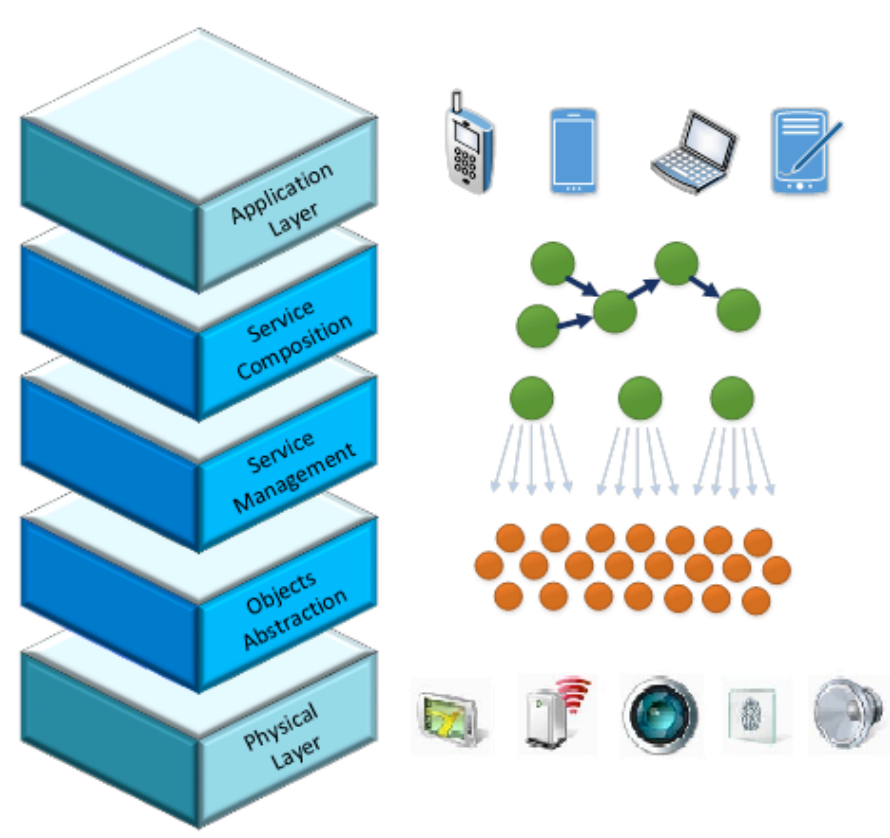

Figure 1: SOA-based architecture for IoT middleware [1].

With the use of SOA, devices can be reused or upgraded individually, leading to several advantages such as extensibility, scalability, and modularity along with the aforementioned interoperability among IoT devices [6, 13].

Because of these advantages of SOA, in [14], the authors presented an energy-centred and Quality of Service (QoS)-aware services selection algorithm (EQSA) for the composition of IoT services. They proposed a model that selects the services by using a lexicographic optimisation strategy and a QoS constraint relaxation technique. The authors of [15] surveyed the recent development of SOA models for IoT and reviewed their fundamental technologies. The authors of [16] proposed a reference architecture based on SOA concepts by integrating the IoT, cloud, and edge technologies with the existing infrastructure. The authors of [17] surveyed the recent development of energy-efficient solutions for wireless sensors networks and reviewed some existing topologies that allow trade-offs between multiple requirements to be achieved for efficient and sustainable sensor networks. The authors of [18] presented a QoS message scheduling algorithm in IoT network-based SOA, which is targeted more toward service provisioning with the idea of service differentiation by classifying the messages into high-priority and best-effort messages. The authors of [19] surveyed the state of QoS methodologies in wireless terrestrial sensor networks to attain the delay and reliability requirements in critical applications. These authors emphasised the main challenges in implementing QoS protocols in Wireless Sensors Networks (WSN) applications. The authors of [20], [21] developed strategies to improve the energy efficiency of Internet of Things, while [22], [23] considered the virtualisation of such networks. Processing the sensor data and the use of analytics based on such big data was surveyed in [24], with [25] using these analytics for effective actuation in the network. Greening these big data networks was introduced and discussed in [26], [27]whereas improving the energy efficiency of the clouds and their interconnecting networks that process the IoT data was evaluated in [28], [29] with the energy efficiency of content sharing optimised in [29] and[30]. The energy efficiency of the networks supporting different services was optimised in [31-39]. Resilience is essential for a range of services, hence [40]and [41] introduce strategies to improve resilience with energy efficiency.
In the present study, we formulated the problem of finding the optimal set of IoT nodes and links to embed BPs into the IoT layer by considering the following three objective functions: i) minimising only the network and processing power consumption, ii) minimising only the mean traffic latency, and iii) minimising a weighted combination of the power consumption and the traffic latency. This problem was formulated using mixed integer linear programming (MILP). We benefit from our track record in energy efficiency and networks virtualization, eg. [42-44]

The rest of this paper is organised as follows: In Section II, we introduce our framework of service embedding in IoT networks. Section III discusses the service embedding evaluation and its results. Finally, Section IV concludes this paper.

## THE FRAMEWORK OF SERVICE EMBEDDING IN IOT NETWORKS

In the smart building setting, many services employ IoT nodes such as: security services employing motion detectors, RFID, display screens and alarms; energy saving services employing motion detection, temperature sensors; fire protection services employing temperature sensors, smoke detectors, water sprinklers and alarms; entertainment services employing noise detectors, and temperature sensors; administration services employing motion detectors, temperature sensors, door actuators, and alarms.
These services and other services can share the same sensing and actuating facilities like sensors for motion, temperature, sound, smoke detectors in addition to the processing modules of the IoT nodes. IoT networks are capable of providing multiple services but they require an efficient architecture that hides such heterogeneity from higher level services and provides interoperability for information exchange with other IoT devices. The SOA enables the abstraction of the IoT node functions and their translated into basic services which in turn can be composed into complex services and exploited by the upper application layer.
We develop a framework to embed service requests into a substrate network of IoT nodes. These requests are implemented following the SOA in the form of a BP. A BP is a virtual topology that consists of virtual nodes and links. The virtual nodes encapsulate the requested processing demand, sensing/actuating functions. The virtual links carry traffic between virtual nodes. The embedding process maps the virtual nodes and virtual links of each BP into nodes and links of the IoT layer.
Each BP is defined as a set of virtual nodes and links. Each virtual node has a function that requires processing and memory. Virtual nodes need to be embedded in a certain geographical zones. Virtual links carry traffic demands between virtual nodes.

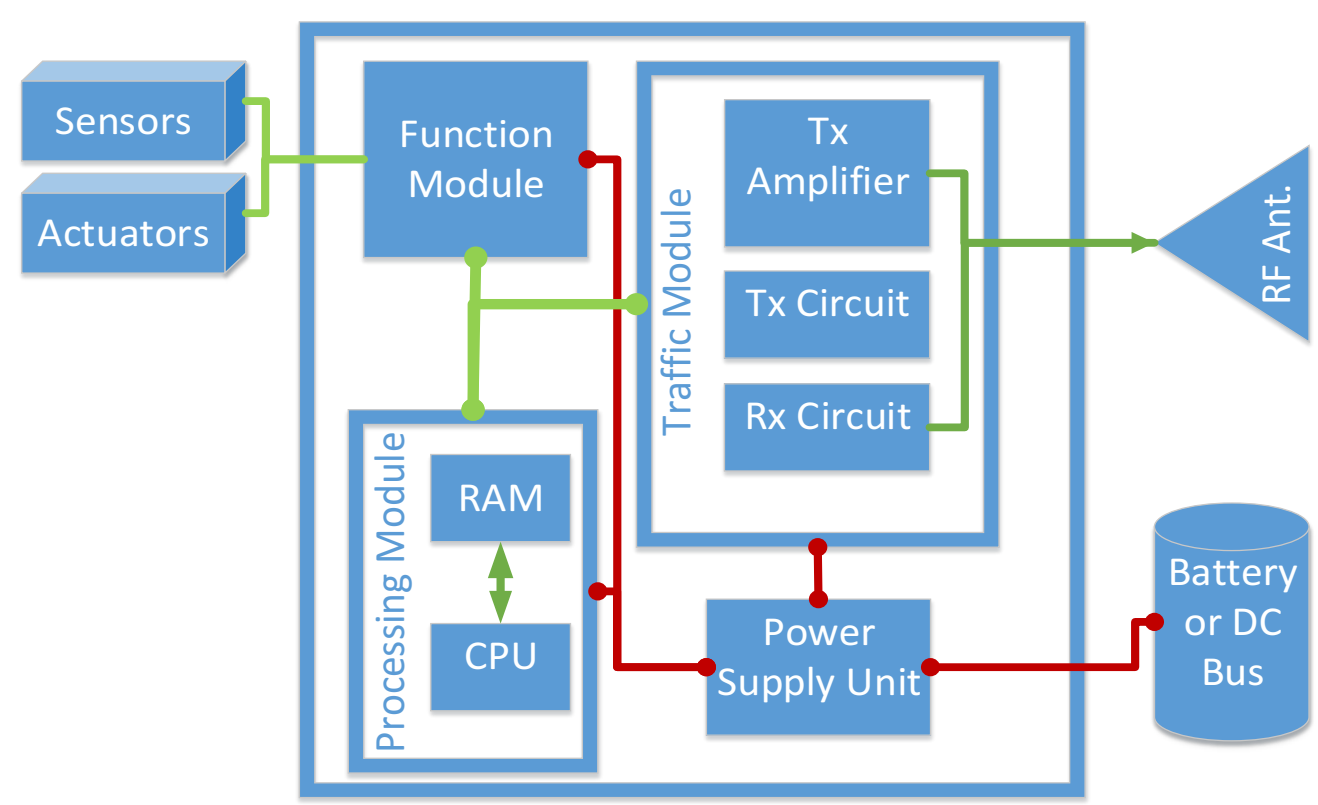

Figure 2: Block diagram of IoT Node.

Each IoT node is characterised by the following modules as shown in figure 2:

- A processing module hosting CPU and RAM.
- A network module hosting a wireless traffic transceiver (Tx/Rx circuit and a Tx power amplifier).
- A function module that provides interfaces to a set of supported sensors and actuators.

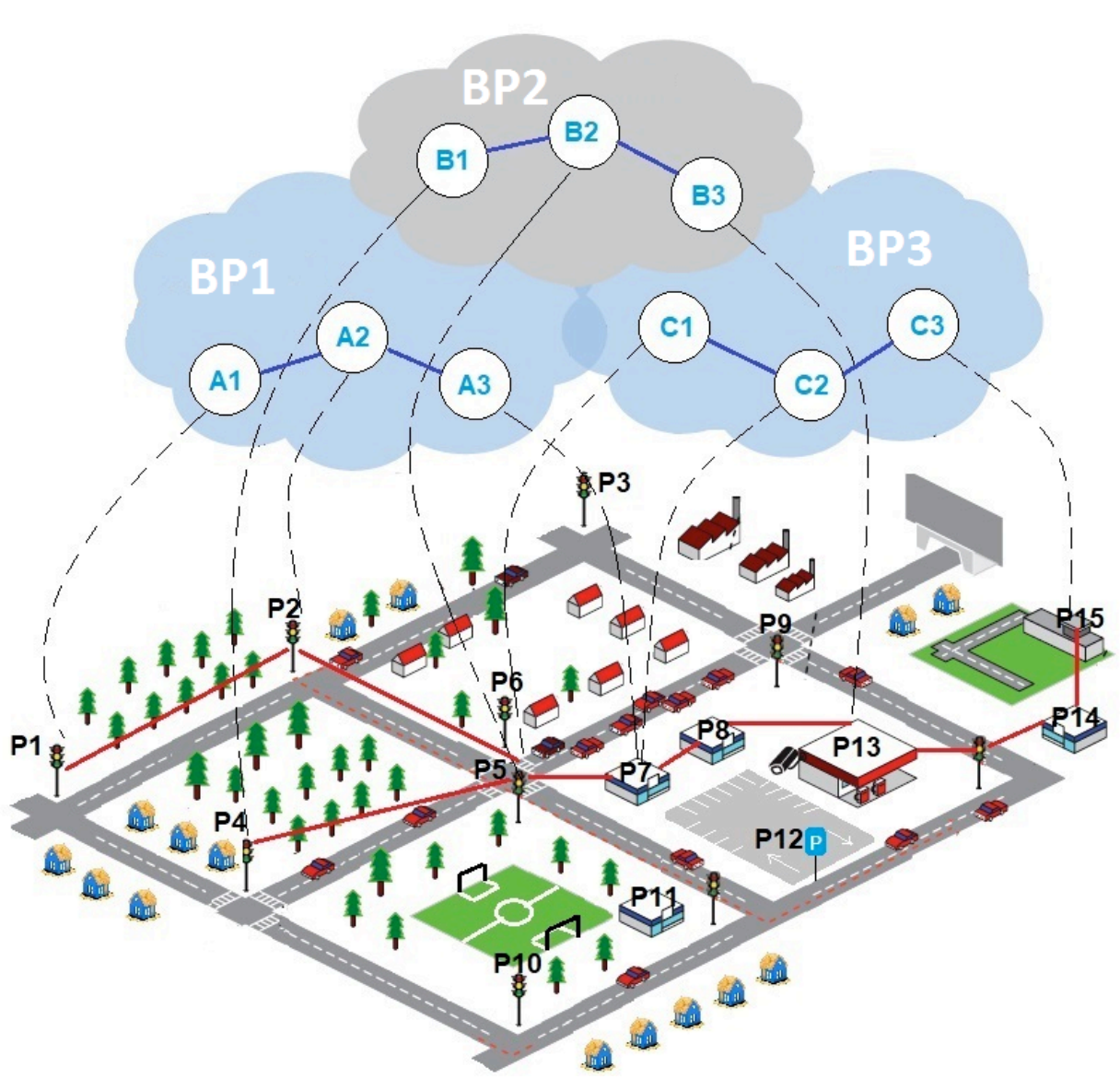

Figure 3: Service embedding layers in IoT networks

Figure 3 gives an example of embedding two BPs. The framework embeds the virtual nodes of BP1 (A1-A2-A3) in the physical IoT nodes (P1-P2-P7), respectively; and chooses the path (P1-P2-P5-P7) to link the embedding IoT nodes. Each virtual node is embedded into an IoT node that satisfies the virtual node's requirements. An IoT node that embeds a certain virtual node of a certain BP can at the same time work as a relay node for the traffic associated with another BP. This is shown in the second embedding example where IoT node P5 which is an embedding node for BP1 and at the same time works as a relay node for the traffic associated with BP2.

We consider a typical IoT setting where the power consumption of IoT nodes is mainly attributed to the processing and network modules while the sensing and actuating modules are externally powered.

As the traffic between IoT nodes is routed via a multi-hop network, we consider the queuing and transmission latency which dominate over the propagation delay as a network performance metric referred to it as traffic mean latency.

To study the power consumption and traffic mean delay resulting from embedding BPs into the IoT network, we formulate the embedding problem as a MILP model considering three different objective functions:

- Minimising the total power consumption.
- Minimising traffic mean latency.
- Minimising both total power consumption and traffic mean latency in a multi-objective manner.

### *A. Framework definitions*

Before we give these objective functions and the constraints imposed on the embedding of BPs, we introduce the sets, parameters and variables used in the formulations:

Sets

| | |
|---|---|
| $B$ | Set of business processes (BPs) in the virtual layer |
| $V$ | Set of virtual nodes in each BP |
| $VN_{ia}$ | Set of neighbours of each virtual node in each BP $(i \in B, a \in V)$ |
| P | Set of IoT nodes in the physical layer |
| $PN_c$ | Set of neighbours of IoT nodes $(c \in P)$ |
| F | Set of functions supported by IoT nodes |
| Z | Set of zones in the IoT physical layer |
| $\lambda$ | Set of arrival rates |
| $W_j$ | Set of traffic mean latency per arrival rate ($j \in \lambda$) in ms per packet |

Parameters

| | |
|---|---|
| $V_{ian}^{FUNC}$ | $V_{ian}^{FUNC} = 1$ If virtual node $a$ in BP $i$ requires the function $n$, $V_{ian}^{FUNC} = 0$ otherwise |
| $V_{iaz}^{ZONE}$ | $V_{iaz}^{ZONE} = 1$ If virtual node $a$ in BP $i$ requires zone $z$, $V_{iaz}^{ZONE} = 0$ otherwise |
| $V_{ia}^{MCU}$ | Processing requirement of the virtual node $a$ in BP $i$ in MHz |
| $V_{ia}^{RAM}$ | Memory requirement of the virtual node $a$ in BP $i$ in kB |
| $V_{iab}^{TRFIC}$ | Traffic demand between the virtual node pair $(a, b)$ in $BP\ i$ in kb/s |
| $P_{cn}^{FUNC}$ | $P_{cn}^{FUNC} = 1$ If IoT node $c$ can provide the function $n$, $P_{cn}^{FUNC} = 0$ otherwise. |
| $P_{cz}^{ZONE}$ | $P_{cz}^{ZONE} = 1$ If the IoT node $c$ is located in zone $z$, $P_{cz}^{ZONE} = 0$ otherwise. |
| $P_c^{MCU}$ | Processing capability of the IoT node $c$ in MHz. |
| $P_c^{RAM}$ | Memory capability of the IoT node $c$ in kB. |
| $P_{ef}^{DIST}$ | Distance between the neighbouring IoT node pair $(e, f)$ in meters. |
| $P_c^{IDLECP}$ | Idle processor power in each IoT node $c$ in mW. |
| $P_c^{MAXCP}$ | Maximum processor power consumption in each IoT node $c$ in mW. |
| $P_c^{IDLETP}$ | Idle network power consumption in each IoT node $c$ in mW. |
| $E_{ef}^{PBT}$ | Energy per bit for each IoT link $(e, f)$ in mW/kbps. |
| M | Large number ($= 10^8$). |
| $P_e^{CAPT}$ | Link capacity for each IoT node $(e)$ in kbps. |
| $F_{ef}^{TR}$ | Transmit amplifier factor for each IoT link $(e, f)$ in mW/kbps/$m^2$. |

Variables

| | |
|---|---|
| $I_{iac}^{NE}$ | $I_{iac}^{NE} = 1$ If virtual node $a$ in BP $i$ has been embedded in IoT node $c$, $I_{iac}^{NE} = 0$ otherwise. |
| $I_{iacn}^{F}$ | $I_{iacn}^{FI} = 1$ If IoT node $c$ has the function $n$ required by virtual node $a$ in BP $i$, $I_{iacn}^{FI} = 0$ otherwise. |
| $I_{iacz}^{Z}$ | $I_{iacz}^{ZI} = 1$ If IoT node $c$ is located in zone $z$ required by virtual node $a$ in BP $i$, $I_{iacz}^{ZI} = 0$ otherwise. |
| $I_{iabcd}^{LE}$ | $I_{iabcd}^{LE} = 1$ If the neighbouring virtual nodes $(a, b)$ in BP $i$ have been embedded in IoT nodes $(c, d)$, $I_{iabcd}^{LE} = 0$ otherwise. |
| $X_{iabcd}^{XOR}$ | Dummy binary variable |
| $R_{cd}^{TRFP}$ | Embedded traffic demand between IoT nodes $(c, d)$ in kbps. |
| $R_{cdef}^{ROUTE}$ | Traffic between IoT nodes $(c, d)$ traversing the |

neighbouring IoT nodes $(e,f)$ in kbps.

$I_{cdef}^{R}$ — $I_{cdef}^{R}=1$ If the traffic demand between IoT nodes $(c,d)$ traverses neighbouring IoT nodes $(e,f)$, $I_{cdef}^{R}=0$ otherwise.

$R_{ef}^{TRFL}$ — Traffic between neighbouring IoT nodes $(e,f)$ in kbps.

$R_{f}^{TRFN}$ — Arrival rate of IoT nodes $(f)$ in kbps.

$LI_{f\,j}^{\text{Lmbda}}$ — Lambda indicator for each IoT node $(f)$ ; $(j) LI_{f\,j}^{\text{Lmbda}}=1$ if the arrival rate is $(j)$, it is 0 otherwise.

$W_{f}^{NODE}$ — Traffic mean latency for each node $(f)$.

$I_{c}^{PM}$ — $I_{c}^{PM}=1$ If the processing module of IoT node $c$ is powered on, $I_{c}^{PM}=0$ otherwise.

$I_{c}^{TM}$ — $I_{c}^{TM}=1$ If the network module of IoT node $c$ is powered on, $I_{c}^{TM}=0$ otherwise.

$TPP$ — Total processing power in the IoT network in mW.

$TNP$ — Total network power in the IoT network in mW.

$TL$ — Total traffic mean latency in ms.

## B. *Framework objective functions*

### 1) *Energy efficient service embedding*

This embedding scenario has an objective function whose goal is to minimise the total power consumption as follows:

**Objective: minimise TNP+TPP** (1)

where TPP is total processing power and given by:

$$TPP=\sum_{c\in P} I_{c}^{PM}\cdot P_{c}^{IDLECP}+\sum_{c\in P}\sum_{i\in B}\sum_{a\in V} I_{iac}^{NE}\cdot P_{c}^{MAXCP}\cdot \quad (2)$$

where $I_{c}^{PM}$ is a binary variable that indicates the activity of the processing module in IoT node c , $P_{c}^{IDLECP}$ is the idle processing power parameter of IoT node c in mW, $I_{iac}^{NE}$ is a binary variable that indicates if a virtual node a in BP *i* has been embedded in IoT node c, $P_{c}^{MAXCP}$ is a parameter that gives the maximum CPU power consumption in each IoT node *c* in mW, $V_{ia}^{MCU}$ is a parameter whose value gives the processing requirement of the virtual node a in BP a in MHz, and $P_{c}^{MCU}$ is a parameter that specifies the processing capability of the IoT node *c* in MHz. The processing power consumption is considered to follow a linear profile versus the load with an idle power consumption. The total traffic power consumption of the network, TNP, and given by:

$$TNP=\sum_{e\in P} I_{e}^{TM}\cdot P_{e}^{IDLETP} \quad (3)$$
$$+2\cdot\sum_{e\in P}\sum_{f\in PN_e} R_{ef}^{TRFIC}\cdot E_{ef}^{PBT}$$
$$+\sum_{e\in P}\sum_{f\in PN_e} R_{ef}^{TRFIC}\cdot (P_{ef}^{DIST})^2\cdot F_{ef}^{TR}$$

where *f* is neighbour IoT node of *e* and is included in $PN_e$, $PN_e$ is the neighbours subset of IoT node *e* , $I_{e}^{TM}$ is a binary variable that indicates the activity of the network module in the IoT node, $P_{e}^{IDLETP}$ is the idle network power parameter of IoT node *e* , $R_{ef}^{TRFIC}$ is a variable that specifies the traffic between neighbouring IoT nodes *e* and *f* in kbps, $E_{ef}^{PBT}$ is a parameter that gives the energy per bit for each IoT link *e*, *f* in mW/kbps, $P_{ef}^{DIST}$ is a parameter that specifies the distance between the neighbouring IoT nodes pair (*e*, *f*) in meters, and $F_{ef}^{TR}$ is the transmit amplifier factor [18] for each IoT link *e*, *f* in mW/kbps/m2.

The network power consumption is a function of the traffic and distance between the source and destination nodes. The network power consumption of each link consists of the idle power, the power consumed per bit by the electronics in the transmitter and the receiver, and the transmitter amplifier power consumption which is calculated based on the radio energy needed based on Frii's free-space equation in our setting (note that higher propagation factors beyond Frii's square law, e.g. cubic or higher, can be considered, and are a straight forward extensions of our equations, but are not considered here) [4, 44, 45].

### 2) *Low latency service embedding*

The second scenario in our framework is concerned with minimising the total traffic mean latency of the service embedding. The framework minimises the traffic mean latency in the IoT network using the following objective function:

**Objective: minimise TL** (4)

where TL $P^{TL}$ is the total traffic mean latency in the network given by:

$$TL=\sum_{f\in P} W_{f}^{NODE} \quad (5)$$

Our network is modelled as an open Jackson network of multiple M/M/1 queues where the utilisation is less than 1 at every queue [46]. For simplicity, we consider each node as an M/M/1 queue. The M/M/1 model refers to a system with a single server, where arrivals are determined by a Poisson process and job service times have an exponential distribution as shown in figure 4.

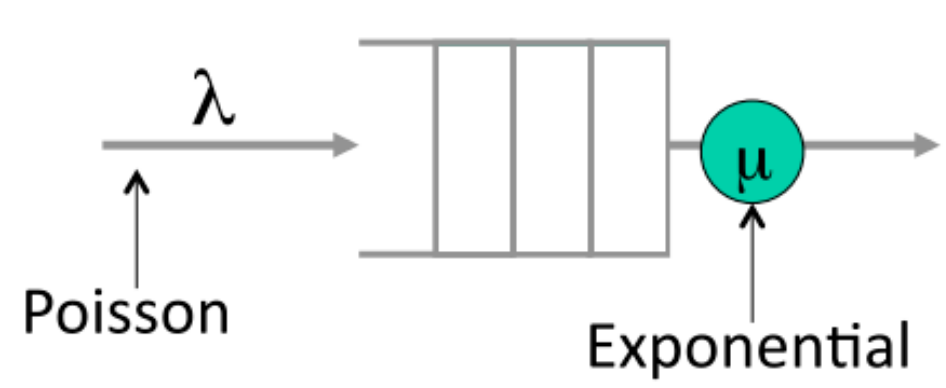

Figure 4: Single server queuing system.

The mean latency is the average time that the packet takes to pass through queue and server, which is given by:

$$W_{f}^{NODE}=\frac{1}{(\mu_{f}^{NODE}-\lambda_{f}^{NODE})} \quad (6)$$

The arrival rate represents the average rate of successful packets transfer to the node through physical links per time unit.

Mathematically, the arrival rate is the summation of data rates delivered to the node in the network.

In our framework, we considered that the service rate $\mu_f^{NODE}$ is fixed for each IoT nodes in the network. The service rate is the transmission rate of the network module. A variable, $\lambda_f^{NODE}$, is created to calculate the summation of packet arrival at each IoT device.

Since we are using linear programming, equation (6) must be converted to a linear format. To facilitate this, we use a lookup table indexed-variable to calculate the traffic mean latency. The lookup table indexed-variables method depends on generating lambda indicator as a binary variable according to the traffic value of $\lambda_f^{NODE}$ for each node. Based on this indicator, the traffic mean latency for IoT nodes is given as the value corresponding to the indicator in the lookup table.

3) *Energy efficient - Low latency service embedding*

In this scenario, we consider a multi-objective MILP model to optimise the service embedding in IoT networks to achieve a trade-off between minimising the power consumption and minimising the traffic mean latency. The objective function is given as:

**Objective: minimise** $\alpha.TL + \beta.TNP + \gamma.TPP$ (7)

where $\alpha$, $\beta$ and $\gamma$ are weight factors with the following units 1/ms, 1/mW, 1/mW respectively used to emphasise the importance of the different components of the objective function.

*C. Framework Constraints*

The framework performs the embedding operation through two parts as follows:

Embedding of virtual nodes

$$\sum_{c \in P} I_{iac}^{NE} = 1 \quad (8)$$

$\forall\, i \in B, \quad \forall\, a \in V$

$$\sum_{a \in V} I_{iac}^{NE} \leq 1 \quad (9)$$

$\forall\, i \in B, \forall c \in P$

Constraint (8) ensures that each virtual node in a BP is embedded in a single IoT node only. Constraint (9) states that each IoT node is not allowed to host more than one virtual node in each BP. This is considered as a coexistence constraint that is not used in all scenarios such as controller node virtualisation.

$$\sum_{i \in B} \sum_{a \in V} I_{iac}^{NE} \geq I_c^{PM} \quad (10)$$

$\forall\, c \in P$

$$\sum_{i \in B} \sum_{a \in V} I_{iac}^{NE} \leq I_c^{PM} \cdot M \quad (11)$$

$\forall\, c \in P$

Constraints (10) and (11) build (include / add) a processing module in IoT node $c$ if that node is chosen for embedding at least one virtual node $a$ in BP $i$ or more, where M is a large enough unitless number to ensure that $I_c^{PMI} = 1$ when $\sum_{i \in B} \sum_{a \in V} P_{iac}^{NE}$ is greater than zero.

$$\sum_{i \in B} \sum_{a \in V} V_{ia}^{MCU} \cdot I_{iac}^{NE} \leq P_c^{MCU} \quad (12)$$

$\forall$ c $\in$ P

$$\sum_{i \in B} \sum_{a \in L} V_{ia}^{RAM} \cdot I_{iac}^{NE} \leq P_c^{RAM} \quad (13)$$

$\forall$ c $\in$ P

Constraints (12) and (13) represent the processing and memory capacity constraints, respectively. They ensure that the embedded processing and memory workloads in an IoT node do not exceed the MCU and memory capacities, respectively.

$$I_{iac}^{NE} \cdot V_{ian}^{FUNC} = I_{iacn}^{F} \quad (14)$$

$\forall$ i $\in$ B, $\forall$ a $\in$ L, $\forall$ c $\in$ P, $\forall$ n $\in$ F

$$P_{cn}^{FUNC} >= I_{iacn}^{F} \quad (15)$$

$\forall$ i $\in$ B, $\forall$ a $\in$ L, $\forall$ c $\in$ P, $\forall$ n $\in$ F

Constraints (14) and (15) ensure that the required function of each virtual node in BP is provided by its hosting IoT node.

$$I_{iac}^{NE} \cdot V_{iaz}^{ZONE} = I_{iacz}^{Z} \quad (16)$$

$\forall i \in B, \forall a \in V, \forall c \in P, \forall z \in Z$

$$P_{cz}^{ZONE} \geq I_{iacz}^{Z} \quad (17)$$

$\forall i \in B, \forall a \in V, \forall c \in P, \forall z \in Z$

Constraints (16) and (17) ensure that the required zone of each virtual node in a BP is matched by the zone of the hosting IoT node.

Embedding of virtual links

$$I_{iac}^{NE} + I_{ibd}^{NE} = X_{iabcd}^{LE} + 2 \cdot I_{iabcd}^{LE} \quad (18)$$

$\forall\, i \in B, \forall a \in V, \forall b \in VN_{ia} : a \neq b, \forall\, c, d \in P : c \neq d$

Constraint (18) ensures that neighbouring virtual nodes $a$ and $b$ of $i$ in $B$ are also connected in the embedding IoT nodes $c$ and $d$. We achieve this by introducing a binary variable $P_{iabcd}^{LE}$ which is only equal to 1 if $I_{iac}^{NE}$ and $I_{ibd}^{NE}$ are exclusively equal to 1 otherwise it is zero, $W_{iabcd}^{LE}$ is an auxiliary variable.

$$\sum_{i \in B} \sum_{a \in L} \sum_{b \in LNB_{ia}} I_{iabcd}^{LE} \cdot V_{iab}^{TRFIC} = R_{cd}^{TRFP} \quad (19)$$

$c, d \in P : c \neq d$

Constraint (19) generates the path's traffic matrix resulting from embedding the virtual nodes $a$ and $b$ into the IoT nodes $c$ and $d$.

$$\sum_{f \in PN_e} R_{cdef}^{ROUTE} - \sum_{f \in PN_e} R_{cdfe}^{ROUTE} \begin{cases} R_{cd}^{TRFP} \\ -R_{cd}^{TRFP} \\ 0 \end{cases} \quad (20)$$

$\forall\, c, d, e \in P : c \neq d \text{ and } e \neq f$

Constraint (20) represents the flow conservation constraint for the traffic flows in the IoT network.

$$\sum_{c \in P} \sum_{d \in P} R_{cdef}^{ROUTE} = R_{ef}^{TRFL} \qquad (21)$$

$$\forall\, e \in P, \forall\, f \in PN_e$$

Constraint (21) estimates link's traffic between the neighbouring IoT nodes $e$ and d.

$$\sum_{f \in PN_e} R_{ef}^{TRFL} \leq P_e^{CAPT} \qquad (22)$$

$$\forall\, e \in P$$

Constraint (22) states that the total traffic flows of the IoT node $e$ should not exceed the node capacity i.e. 250 kbps.

$$R_{cdef}^{ROUTE} \geq I_{cdef}^{R} \qquad (23)$$

$$\forall\, c, d, e \in P, \forall f \in PN_e : c \neq d, e \neq f$$

$$R_{cdef}^{ROUTE} \leq I_{cdef}^{R} \cdot M \qquad (24)$$

$$\forall\, c, d, e \in P, \forall f \in PN_e : c \neq d, e \neq f$$

The constraints (23) and (24) build a path between the embedding IoT nodes $c$ and $d$ through the neighbouring IoT nodes $e$ and $f$, where $I_{cdef}^{RI} = 1$ if there is a traffic path between the IoT nodes $c$ and $d$ that passes through the neighbouring IoT nodes $e$ and $f$, where $M$ is a large enough unitless number which ensure that $I_{cdef}^{R} = 1$ when $R_{cdef}^{ROUTE}$ is greater than zero.

$$\sum_{f \in PN_e} I_{cdef}^{R} \leq 1 \qquad (25)$$

$$\forall c \in P, \forall d \in P, \forall e \in P$$

Constraint (25) ensures that traffic splitting is prevented for each path between the embedding IoT nodes $c$ and $d$, such that the maximum number of physical links between neighbouring IoT nodes $e$ and $f$ is one.

$$\sum_{c \in P} \sum_{d \in P} \sum_{f \in PNB_e} I_{cdef}^{R} \geq I_e^{TM} \qquad (26)$$

$$\forall\, e \in P$$

$$\sum_{c \in P} \sum_{d \in P} \sum_{f \in PNB_e} I_{cdef}^{R} \leq I_e^{TM} \cdot M \qquad (27)$$

$$\forall\, e \in P$$

Constraints (26) and (27) build a network module in IoT node $e$ if that IoT node is chosen to send/receive traffic at least for one link or more, where $M$ is a large enough unitless number to ensure that $I_e^{TMI} = 1$ when $\sum_{c \in P} \sum_{d \in P} \sum_{f \in PNB_e} I_{cdef}^{R}$ is greater than zero.

$$\sum_{e \in PN_f} R_{ef}^{TRFL} = R_f^{TRFN} \qquad (28)$$

$$\forall\, f \in P : e \neq f$$

Constraint (28) estimates the arrival traffic for each IoT node.

$$\sum_{j \in J} LI_{f\,j}^{LMBDA} \cdot j = R_f^{TRFN} \qquad (29)$$

$$\forall\, f \in P : e \neq f$$

Constraint (29) is an arrival rate indicator of arrival rate $j$ for each IoT node $f$

$$\sum_{j \in J} LI_{f\,j}^{LMBDA} \leq 1 \qquad (30)$$

$$\forall\, f \in P$$

Constraint (30) ensures that each IoT node has no more than one arrival rate indicator.

$$\sum_{j \in J} W_j^{LIMDA} \cdot LI_{f\,j}^{LMBDA} = W_f^{NODE} \qquad (31)$$

$$\forall f \in P$$

Constrain (31) estimates the mean traffic latency for each IoT ($f$). The MILP optimisation model was solved using CPLEX running on personal computer with processor core i5 -3.2 GHz and 16 GB RAM and on the university Polaris servers using 24 cores and 128GB RAM.

## [redacted]Results and Evaluations

To evaluate the performance of the proposed model and heuristic, we consider a smart building scheme (for example in an enterprise campus) where the physical layer is composed of 30 IoT nodes connected by 89 bidirectional wireless links. These IoT nodes are distributed across an area 500 m x 500 m and can carry various functions with the following assumptions:

- There is a set of 9 distinct functions, 4 sensing functions, one control function and 4 actuating functions. Each IoT node can provide 2 sensing functions, 2 actuating functions, and one controlling function (present only in one type of processor). The virtual node of each BP requests one function only.
- There is a set of five geographical zones that represent the sub-sections of the smart building (e.g. departments or sections in the enterprise campus). Each zone is equipped with six IoT nodes. All the functions and processor types exist in each zone. The virtual node requests an embedding location in one of these five zones.
- The IoT nodes processing capability is uniformly distributed among five processing capacities (8, 16, 16, 25, 25, 48 MHz) representing microcontrollers as shown in Table 1. Each virtual node has a specific processing demand that varies between 4 and 30 MHz.
- Each IoT node contains wireless transceiver modules [47]. The network modules used are low cost, low power, and are compatible with the ZigBee protocol stack for IoT networks [44]. The traffic demands of the virtual links vary from 50 to 200 packets per second with a packet size of 1 kb.

We study the embedding of 12 BPs arriving sequentially, two at a time. Each BP has three virtual nodes (sensor, controller and actuator) connected sequentially. The sensor is connected to the controller and the controller is connected to the actuator. The sensor virtual node requests a specific sensing function, the control virtual node requires processing capacity and the actuator virtual node requests a specific actuating function. The sensor and actuator virtual nodes of a BP need to be embedded in a specific zone while the controller virtual node can be embedded into any geographical zone.

| MCU Type | MCU CLK | RAM | Idle Power | Max. Power |
|---|---|---|---|---|
| MSP430F1 | 8 MHz | 64 kB | 1 mW | 8 mW |
| MSP430FR5 | 16 MHz | 64 kB | 1 mW | 14 mW |
| MSP430FR6 | 16 MHz | 128 kB | 1 mW | 20 mW |
| MSP430F5 | 25 MHz | 512 kB | 1 mW | 14 mW |
| MSP432P4 | 48 MHz | 256 kB | 1 mW | 16 mW |

Table 1: Processing modules power specifications and power consumption in active mode

We evaluate the power consumption and traffic mean latency resulting from embedding the BPs using the MILP model considering the three objective functions.

### A. *Energy efficient service embedding*

In this section, we evaluate the results of embedding BPs in terms of power consumption and traffic mean latency under three scenarios. In the first scenario, referred to as energy-latency unaware service embedding (ELUSE), BPs are embedded in physical nodes and links that satisfy their requirements where objective function's goal is to ensure that all requests for embedding are met.

In the second and third scenarios, the objective is to minimise the total power consumption. However, in the second scenario, referred to as re-provisioning, each time a new BPs arrives, previously embedded BPs are re-embedded while in the third scenario, referred to as sequential embedding, arriving BPs are embedded without interrupting the existing BPs. We also study the coexistence constraints of the embedding and their effects on the results of the energy efficient service embedding.

#### 1) *Service embedding on same geographical zone*

In this subsection, we considered that the sensor and actuator nodes of a BP need to be embedded in the same specific geographical zone. We also study embedding BPs with and without coexistence constraints. Under coexistence constrains, the virtual nodes of the BP cannot coexist in the same IoT node. The goal here is to improve the resilience of the BPs under single node failure.

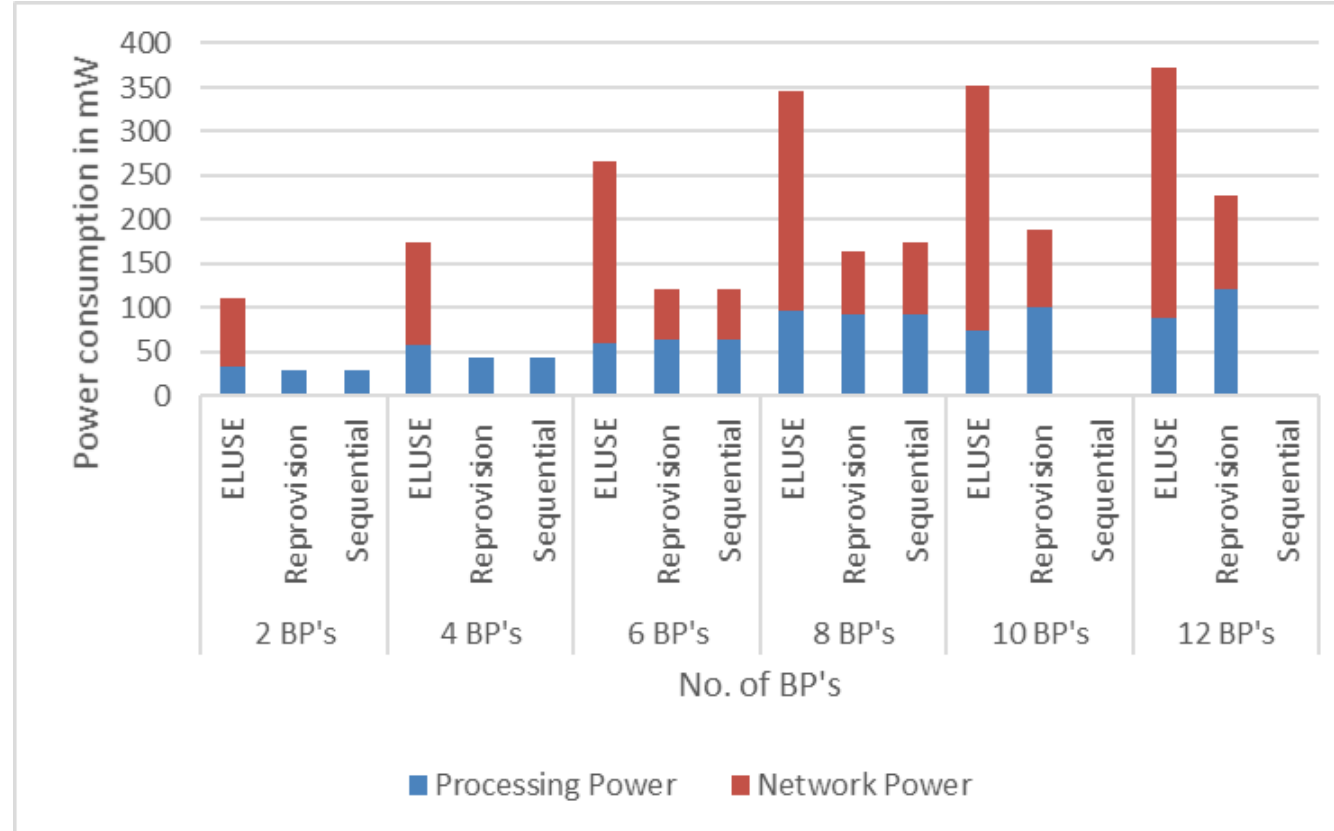

Figure 5: Power consumption of energy efficient service embedding in same zone without coexistence constraint.

Figure 5 shows the total power consumption of embedding BPs in which the sensing and actuating nodes are to be embedded in the same zone. The results show that the energy efficient re-provisioning embedding scenario resulted in saving an average of 63% of the power consumption compared to the ELUSE scenario. Under energy efficient embedding, fewer IoT nodes and links are activated to embed BPs compared to embedding under the ELUSE scenario. As no coexistence constraints apply, all the virtual nodes of a BP can be embedded in a single IoT node confining the virtual links traffic within this node and reducing the number of activated IoT nodes. The saving achieved by the energy efficient embedding decreases to 58% under the sequential scenario as the sequential approach builds on existing embedding decisions that become suboptimal with the arrival of new BPs. The optimal use of resources under the re-provisioning scenario resources resulted in successfully embedding 12 BPs while only 8 BPs were successfully embedded under sequential embedding. Note that the power savings decrease as the number of embedded BPs increases. This is because the higher the load on the network the fewer the possible embedding solutions therefore narrowing the gap between energy efficient embedding and ELUSE.

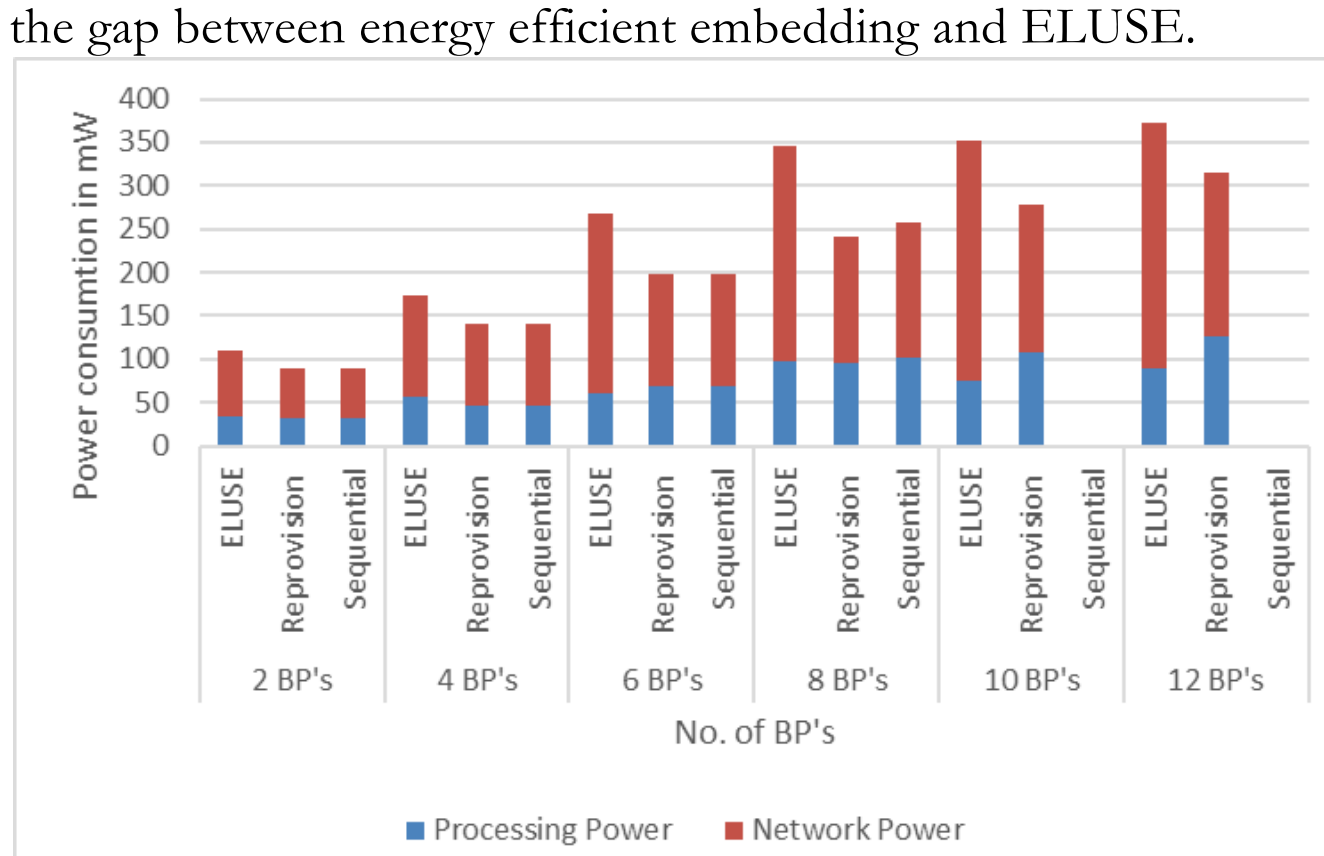

Figure 6: Power consumption of energy efficient service embedding in the same zone with coexistence constraint.

Figure 6 shows the power consumption of embedding BPs in the same zone under coexistence constraints. The coexistence constraints reduce the power savings achieved by the energy efficient embedding scenarios to 36% and 29% for re-provisioning and sequential embedding, respectively. This reduction in power savings is due to the need to activate more IoT nodes to meet the coexistence requirements and the traffic between these nodes.

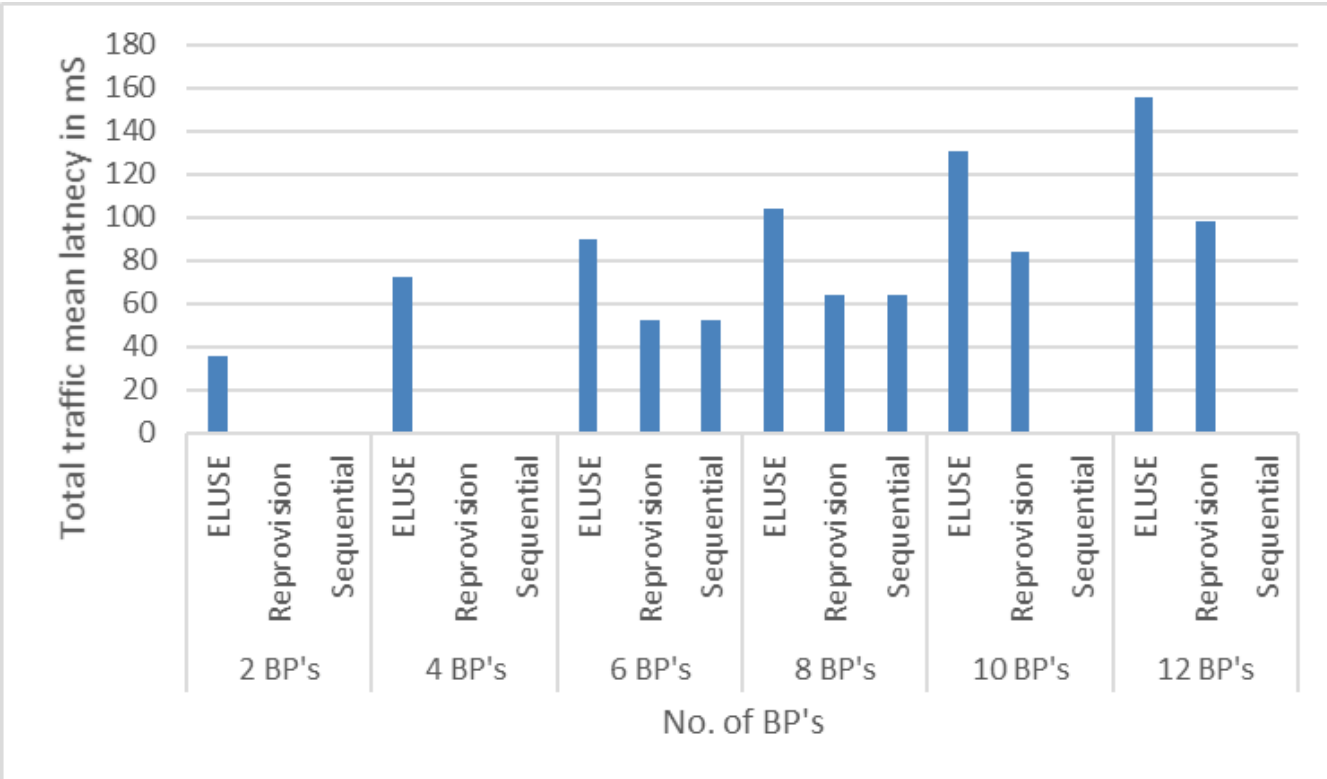

Figure 7: Average traffic mean latency of energy efficient service embedding in same zone without coexistence constraint.

The results in figure 7 display the average traffic mean latency resulting from embedding BPs without coexistence constraint. The re-provisioning embedding and the sequential embedding have reduced the average traffic mean latency by 62% and 60% respectively compared with ELUSE scenario. This is because energy efficient embedding selects routes of minimum hops and consequently lower traffic mean latency compared to random routing in ELUSE. However, energy efficient embedding does not produce the minimum traffic mean latency as energy efficient embedding tries to highly utilise the activated IoT nodes resulting in high traffic mean latency in these nodes.

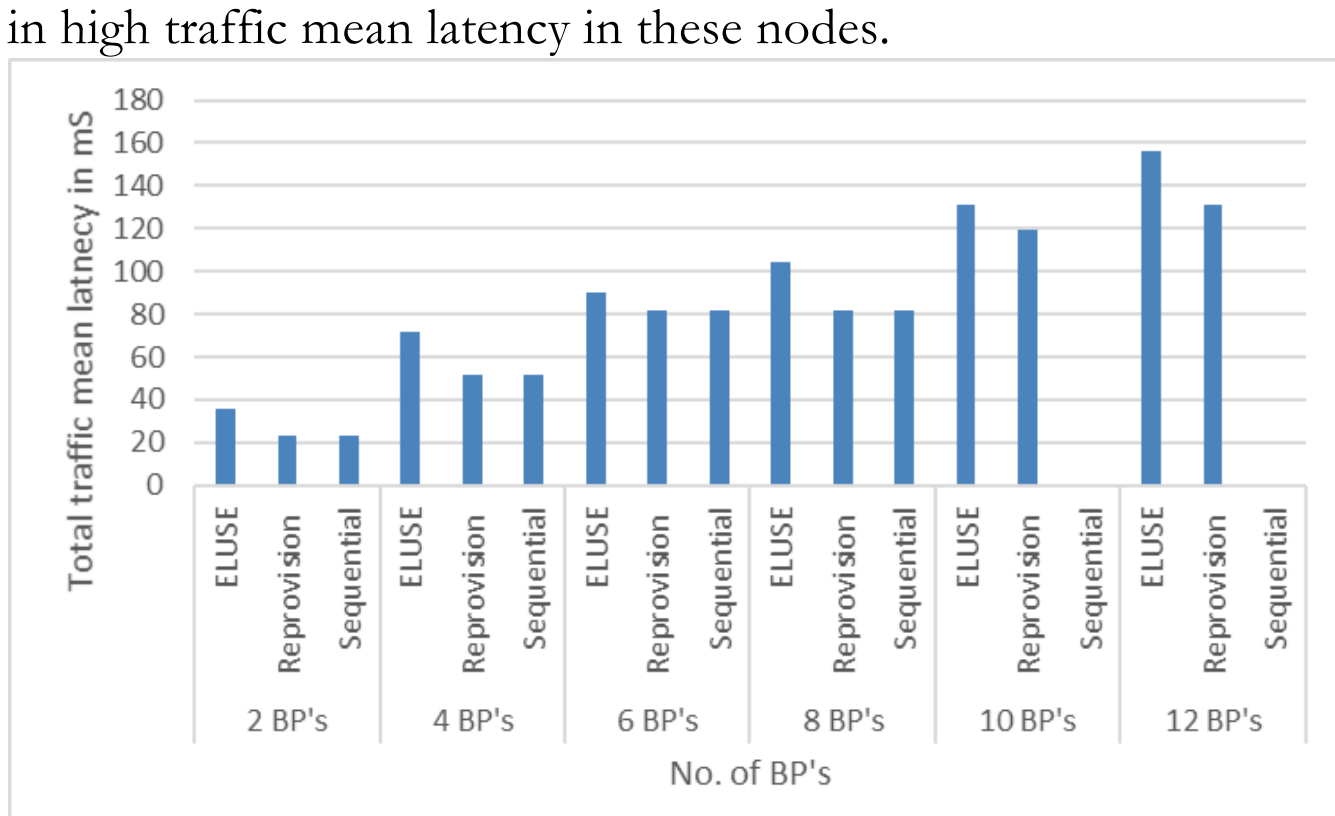

Figure 8: Average traffic mean latency of energy efficient service embedding in same zone with coexistence constraint

Similar trends to those in figure 7 are observed in figure 8 for the average traffic mean latency resulting from embedding with coexistence constraints. The results show that the re-provisional embedding and the sequential embedding have reduced the average traffic mean latency by 27% compared with the ELUSE scenario. Comparing figure 7 and 8 shows that embedding BP on the same zone with coexistence constraint results in higher traffic mean latency compared to embedding without coexistence constraint. This is because without the coexistence constraint, the traffic of a BP can experience no traffic latency by embedding all the virtual nodes of the BP in a single IoT node.

2) *Service embedding across geographical zone*

The previous results evaluated the power consumption and mean latency of embedding BPs where the sensor and actuator nodes need to be embedded in the same geographical zone. In this section we examine embedding BPs that require the sensor and actuator nodes to be embedded in distinct geographical zones. We study also the performance with and without coexistence constraints on the controller node. Under coexistence constrains, the controller cannot coexist in the same IoT node with the sensor or actuator node.

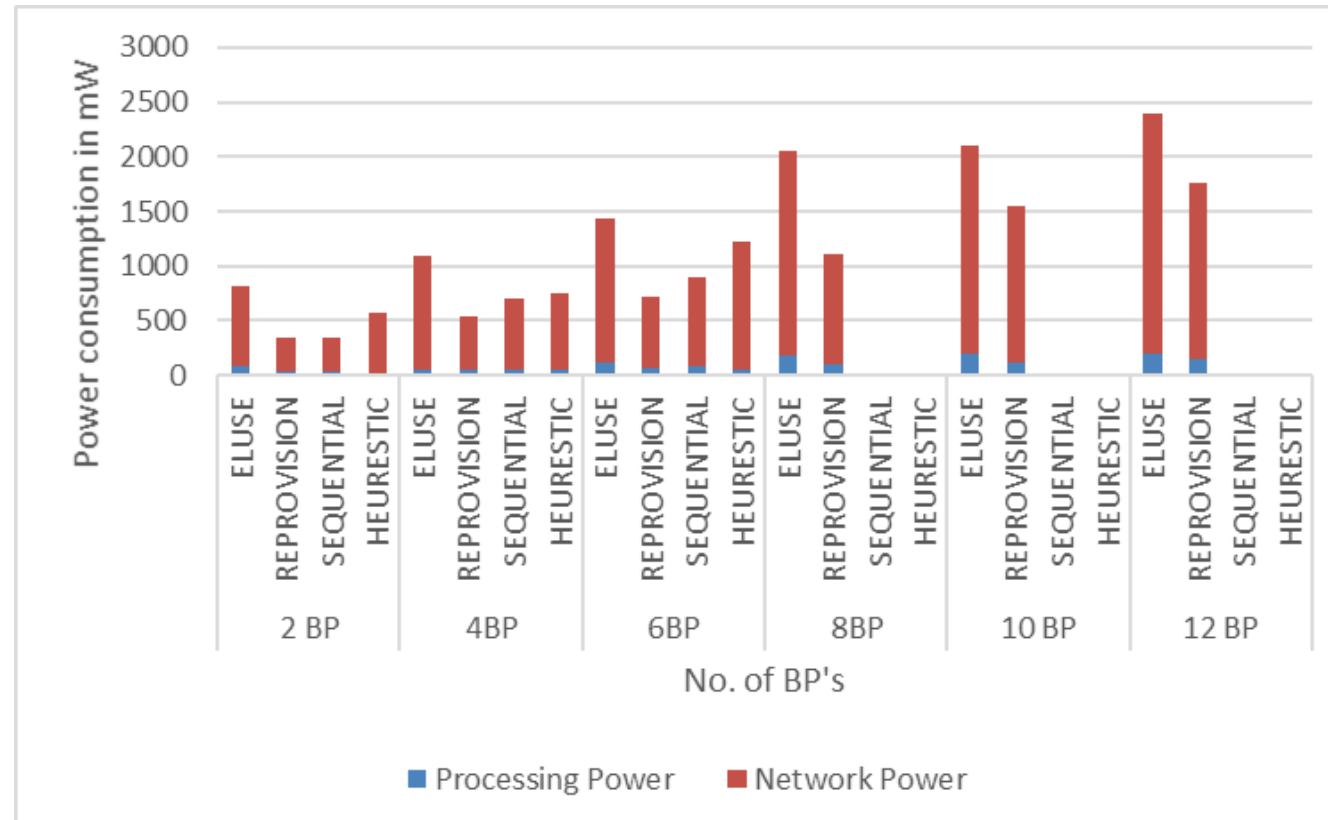

Figure 9: Power consumption of energy efficient service embedding across different zones without coexistence constraint.

Figure 9 displays the power consumption of embedding BPs across different geographical zones without coexistence constraint. The power savings achieved by energy efficient embedding under the re-provisioning scenario and the sequential scenario when embedding across different zones are lower than those achieved for same zone embedding in figure 6. This is because energy efficient embedding in the distinct zones cannot select to embed the sensor and actuator in the same node although coexistence constraints do not apply. The power savings achieved by the energy efficient embedding scenarios are 42% and 22% for re-provisioning and sequential scenarios, respectively.

The less efficient use of resources in embedding across zones reduces the number of BPs that can be embedded under the sequential scenario to 6 BPs, while the re-provisioning embedding still succeeds to embed all the 12 BPs.

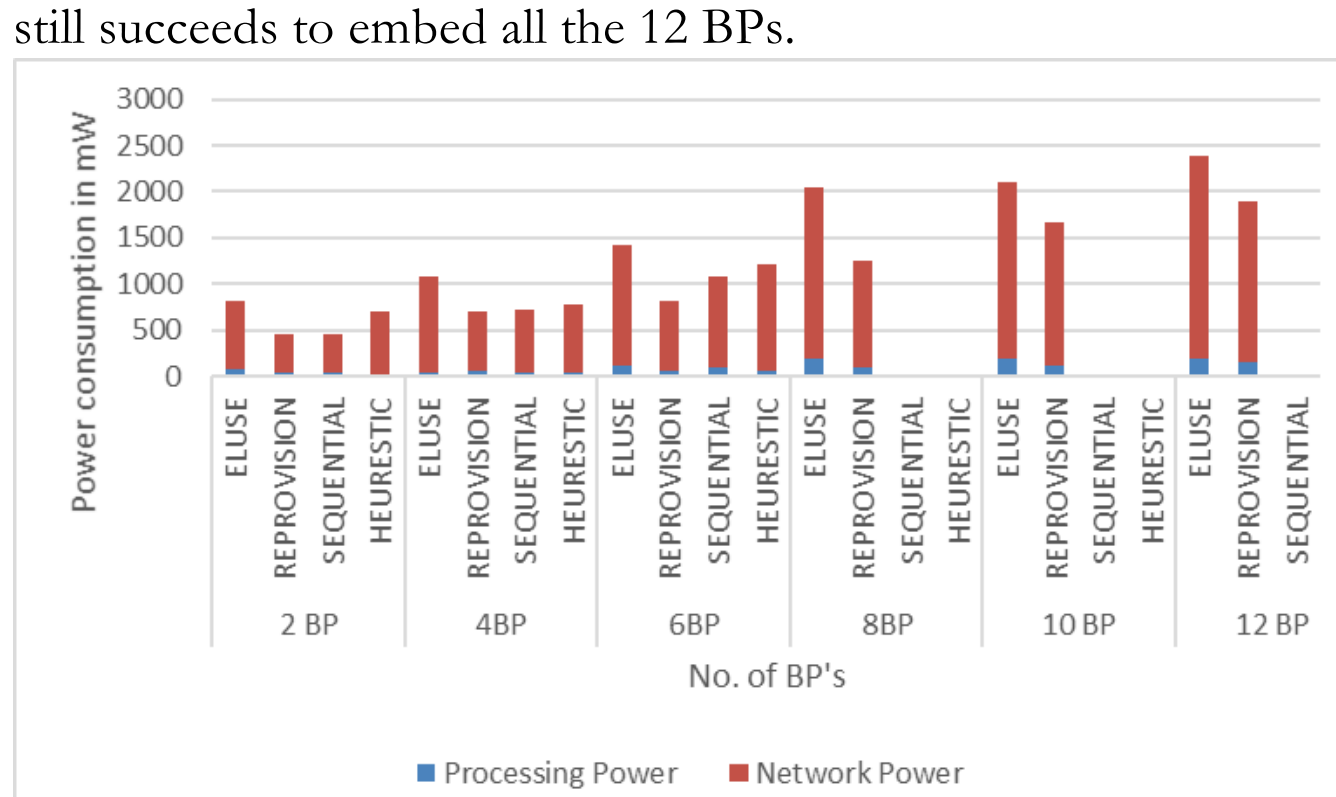

Figure 10: Power consumption of energy efficient service embedding across different zones with coexistence constraint.

Figure 10 summarises the power consumption results when embedding BP's into the physical IoT network with the coexistence constraint. The power savings achieved by the energy efficient embedding scenarios are reduced to 34% and 17% for re-provisioning and sequential cases, respectively. This reduction is due to embedding of virtual nodes of a BP in different IoT nodes as explained above.

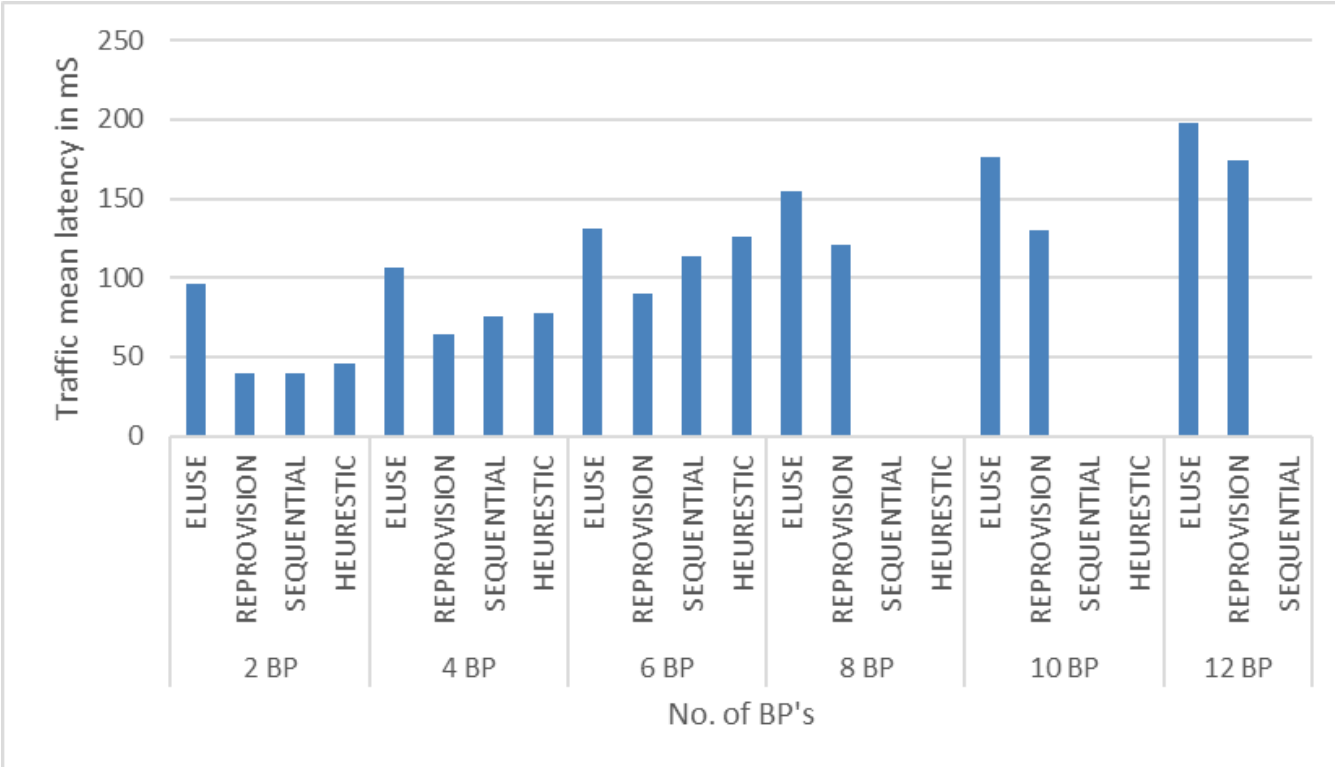

Figure 11: Average latency of energy efficient service embedding across different zones without coexistence constraint.

The traffic mean latency resulting from embedding BPs across distinct zones without coexistence constraints are shown in figure 11. The re-provisioning and sequential embedding have reduced the average traffic mean latency by 32% and 15% compared with ELUSE scenario.

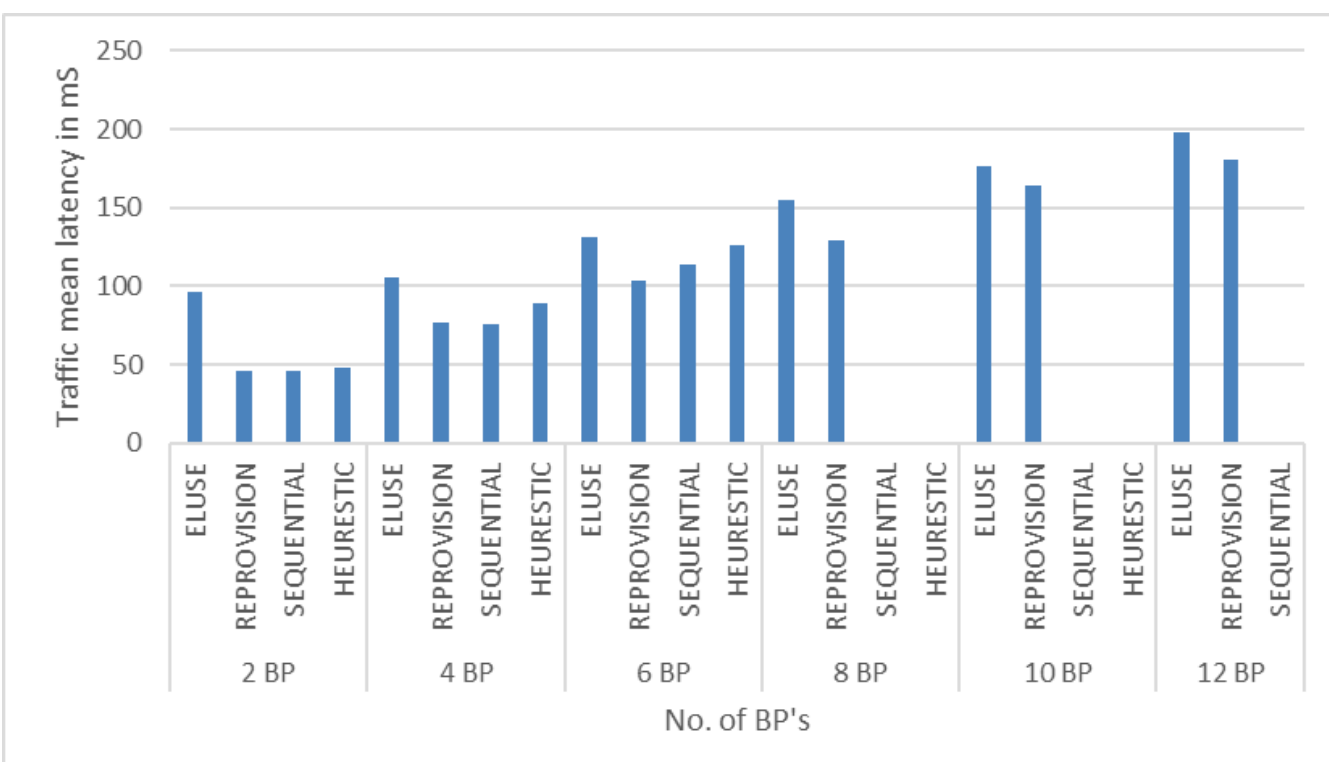

Figure 12: Average latency of energy efficient service embedding across different zones with coexistence constraint.

Figure 12 displays the traffic mean latency resulting from embedding BPs across distinct zones without coexistence constraints. The re-provisioning and sequential embedding have reduced the average traffic mean latency to 22% and 13% compared with ELUSE scenario.

### B. *Low latency service embedding in IoT networks*

In this subsection, we evaluate the low traffic mean latency when embedding of BPs across different zones with and without the coexistence constraint and also assess the power consumption.

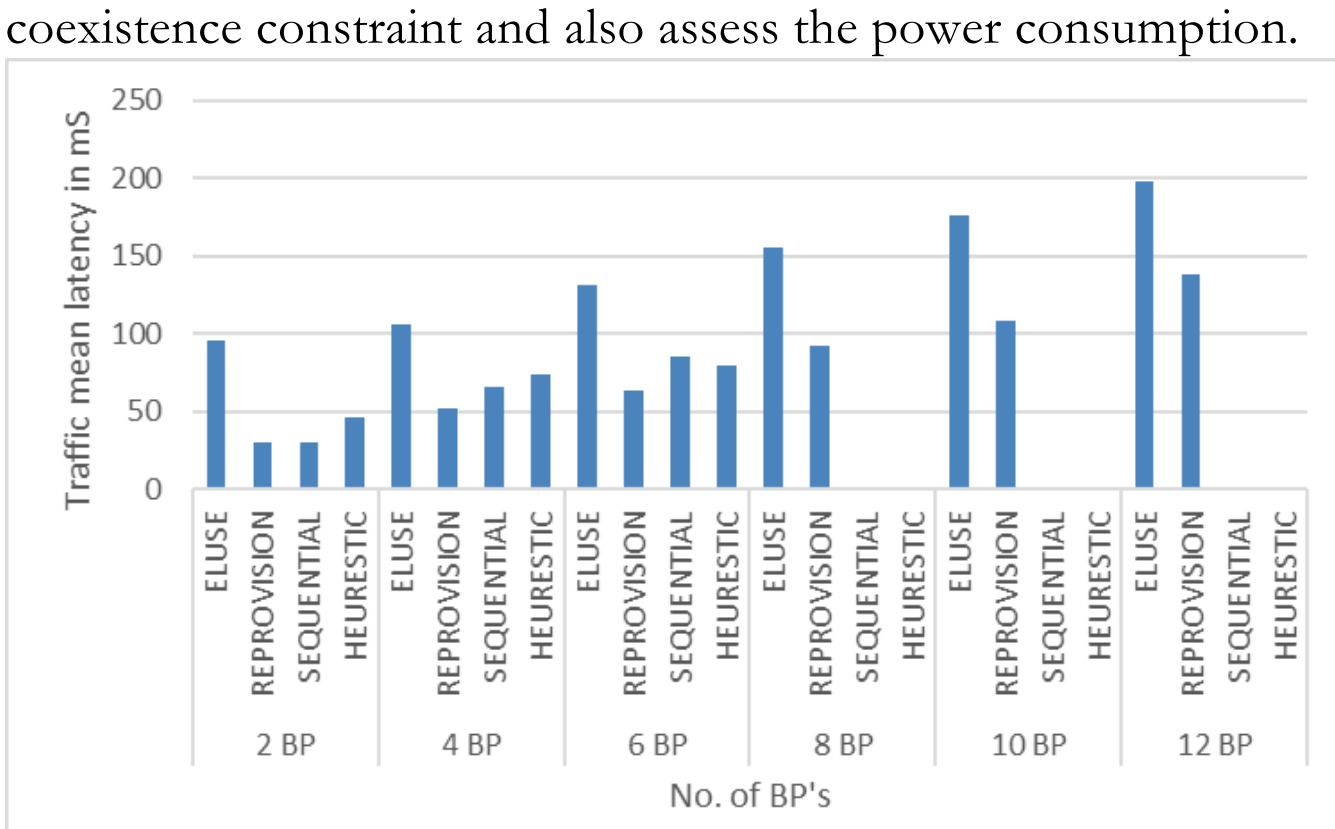

Figure 13: Average traffic mean latency of low latency service embedding across different zones without coexistence constraint.

Figure 13 shows that the re-provisioning low latency embedding resulted in reducing the traffic latency by an average of 47% compared to the ELUSE scenario. The low latency embedding model optimises the selection of IoT nodes and distributes the traffic so that the arrival rate at nodes and consequently the traffic latency is minimised. Under energy efficient embedding, fewer IoT nodes and links are activated to embed BPs compared to embedding under the ELUSE scenario.

The traffic latency reduction achieved by the energy efficient embedding decreases to 20% under the sequential scenario as the sequential approach builds on existing embedding decisions as explained in Section (A). The optimal use of resources under the re-provisioning scenario resulted in successfully embedding 12 BPs while only 6 BPs were successfully embedded under sequential embedding.

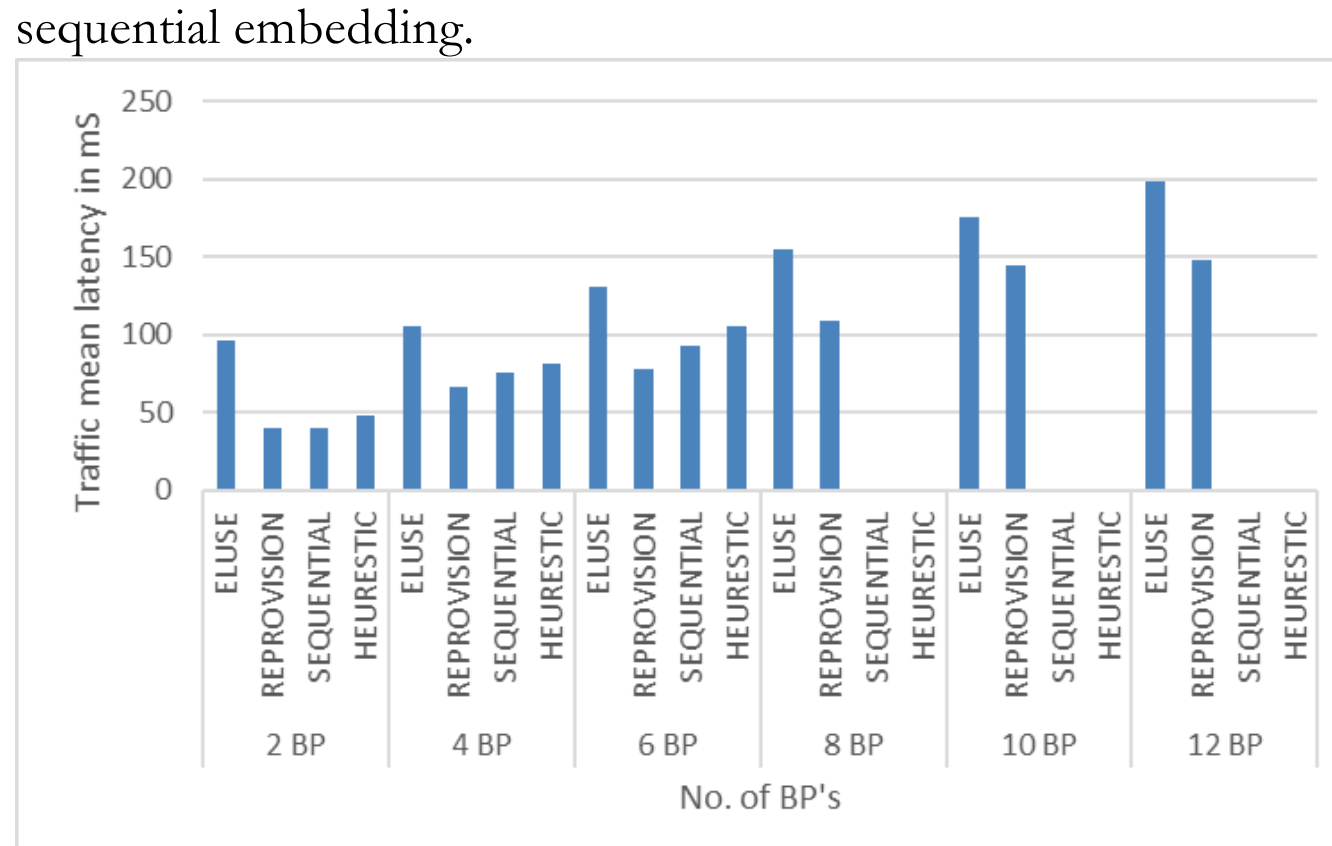

Figure 14: Average traffic mean latency of low latency service embedding across different zones with coexistence constraint.

Figure 14 displays the traffic mean latency of low latency BPs embedding across different zones with the coexistence constraint. Adding the coexistence constraint reduced the traffic latency achieved by the re-provisioning and sequential embedding to 34% and 19%, respectively compared to the ELUSE scenario as more traffic traverses the network due to the fact that multiple virtual nodes of the same BP cannot coexist on the same IoT node.

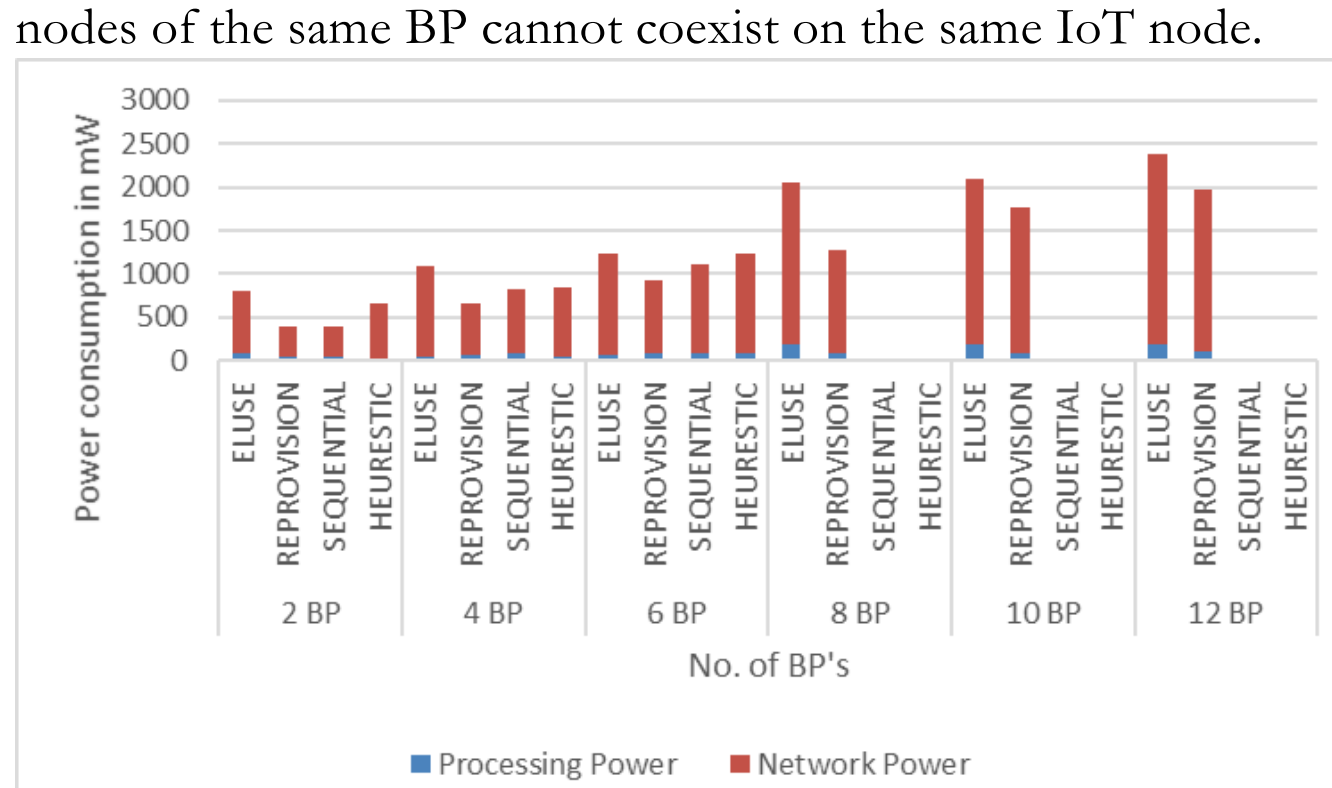

Figure 15: Power consumption of low latency service embedding across distinct zones without coexistence constraint.

The results in figure 15 show the power consumption resulting from low latency embedding across distinct zones without coexistence constraint. Distributing the traffic to reduce the delay increased the power consumption by 28% compared to the energy efficient re-provisioning embedding in figure 9 as more nodes are activated. However, compared to the ELUSE scenario the power consumption is reduced by 18% and 10% under low latency re-provisioning and low latency sequential embedding, respectively.

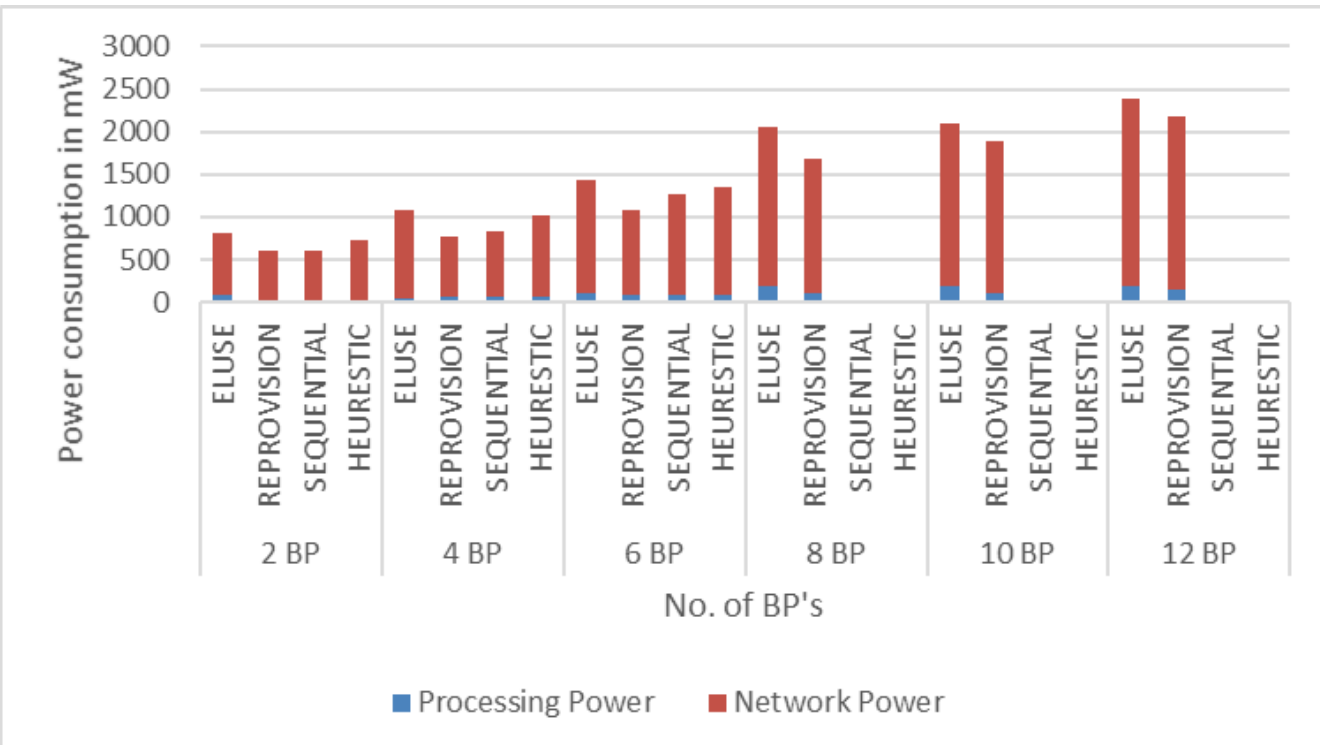

Figure 16: Power consumption of low latency service embedding across distinct zones with coexistence constraint.

Under the coexistence constraint in figure 16, the increase in power consumption resulting from low latency embedding compared to the energy efficient embedding increased the power consumption by 20% compared to the energy efficient re-provisioning embedding in figure 10. However, compared to the ELUSE scenario the power consumption is reduced by 14% under low latency re-provisioning and sequential embedding.

## C. *Energy efficient-Low latency service embedding in IoT networks*

Minimum power consumption is achieved by consolidating the embedding of virtual nodes in as few as possible energy efficient IoT nodes. On the other hand, minimum traffic mean latency is achieved by distributing the traffic into multiple paths to reduce the arrival rate at the individual IoT nodes. As explained in the previous section, the trade-off between minimising the power consumption and minimising the traffic mean latency is achieved through a multi-objective MILP model. We define a metric referred to as "embedding optimality" to compare the performance of the multi-objective embedding to single objective embedding. The embedding optimality is defined as follows:

$$Optimality^{QoS} = \frac{Optimal^{QoS}_{Multi-objective}}{Optimal^{QoS}_{Single-objective}} \quad (33)$$

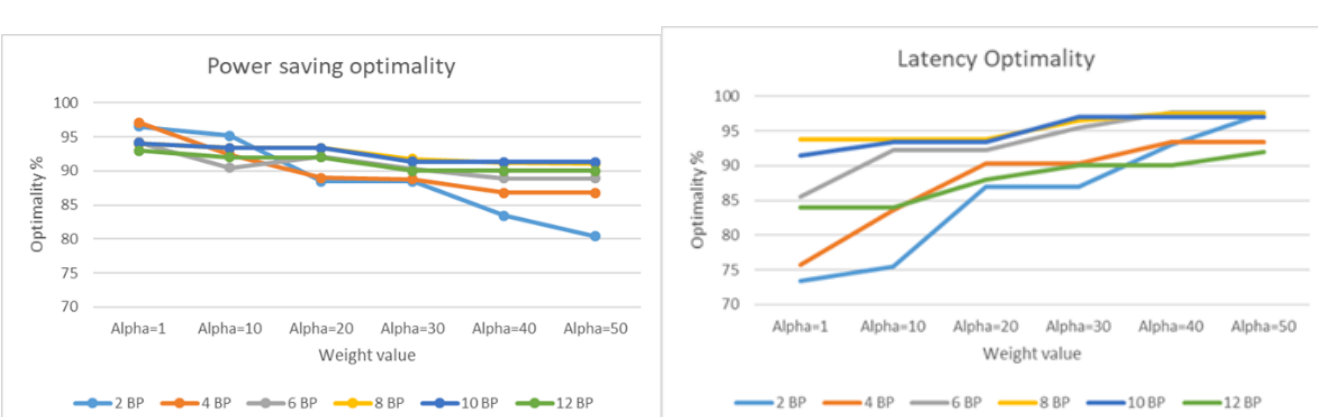

(a) (b)

Figure 17: Optimality of (a) power saving and (b) traffic mean latency of embedding in distinct zones with coexistence constraint.

Figure 17 displays the power saving (figure 17(a)) and traffic mean latency (figure 17(b)) average optimality of energy efficient–low latency service embedding scenario across distinct zones with coexistence constraint under $\alpha = 30$ , $\beta = 1$ and $\gamma = 1$ in the multi-objective function (equation (7)). Note that the numerical value of power consumption and traffic latency are comparable, therefore the weight $\alpha$ is used to prioritise traffic latency, while the other two weights in equation (7) are set to one. We obtain equal optimality for power savings and mean traffic latency of 91% at $\alpha$=30, i.e. this is the weight needed to achieve the trade-off.

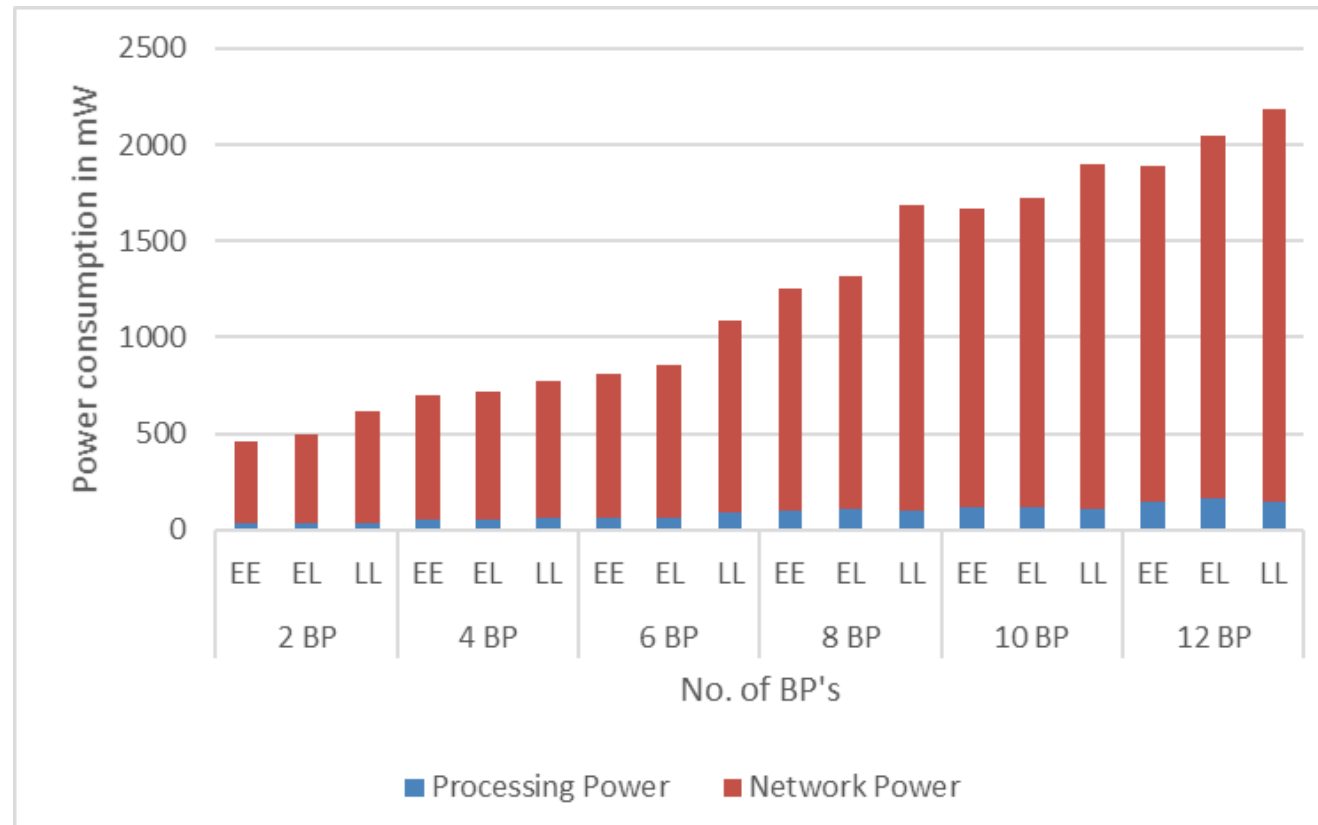

Figure 18: Power consumption of embedding in distinct zones with coexistence constraint.

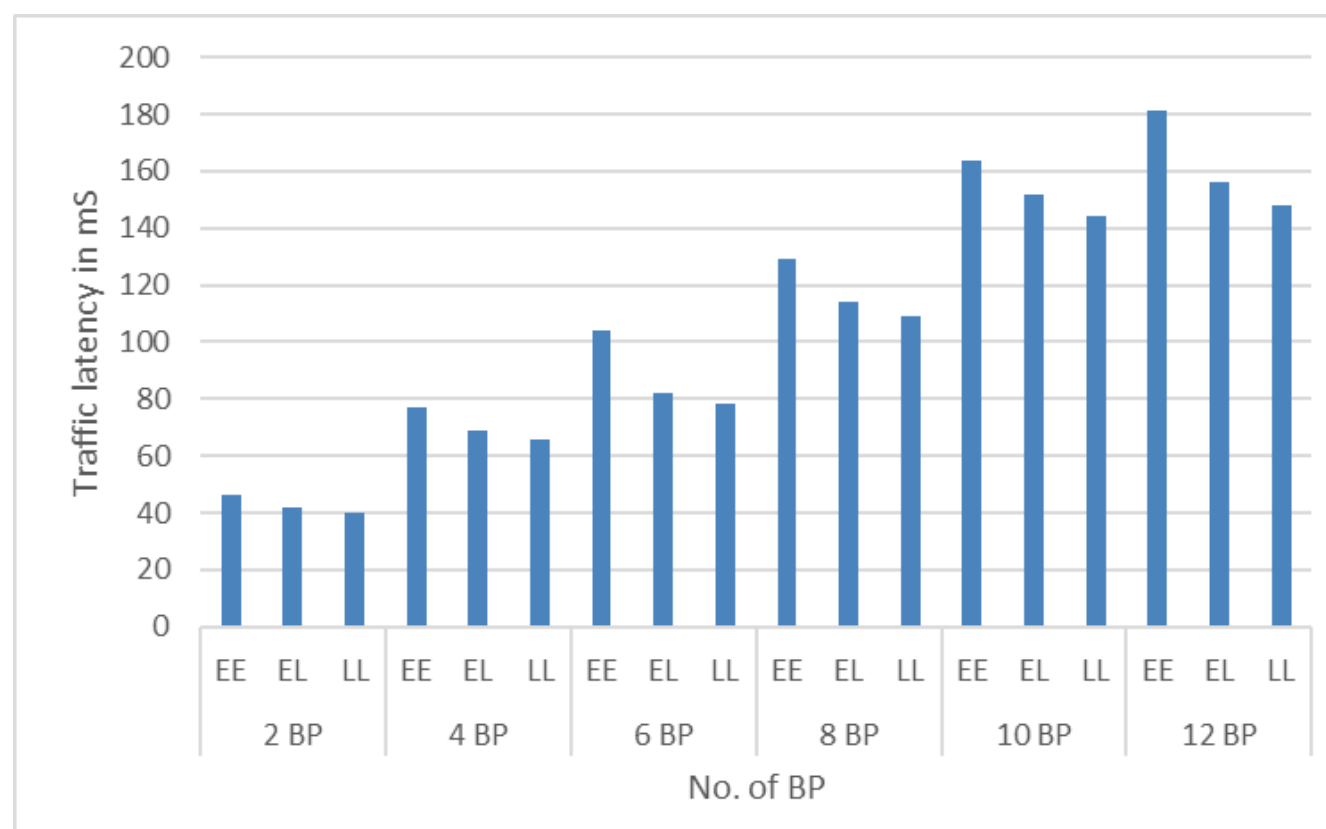

Figure 19: Average traffic mean latency of embedding in distinct zones with coexistence constraint.

Figure 18 and 19 compare the power consumption and delay, respectively of the energy efficient–low latency service embedding scenario with $\alpha = 30$ to those of the energy efficient service embedding and low latency service embedding scenarios. Note that the low latency scenario increases the power consumption by 20% compared to the energy efficient scenario (figure 18) and the energy efficient scenario increases the traffic mean delay by 22% compared to the low latency scenario (figure 19).

## D. *Real time energy efficient service embedding heuristic*

The flowchart of the RESE heuristic is shown in figure 20. The input to the heuristic is the IoT network topology and the BPs. The heuristic starts by sorting the IoT nodes according to the processing power efficiency in descending order and the BPs according to the processing demand of the controller node in ascending order.

The heuristic picks a BP form the ordered list and embeds its nodes one by one considering the IoT node with the highest energy efficiency that satisfies the embedding requirements in terms of function, zone and coexistence. By doing so the heuristic tries to consolidate virtual nodes into the most energy efficient IoT node that meets its demand before activating another IoT node. The available processing capacity of the IoT nodes is updated after the embedding of a virtual node and another virtual node of the BP is selected to be embedded. After embedding all the virtual nodes of a BP, the traffic between the virtual nodes is routed based on shortest path routing [48]. This process is repeated for all BPs and the total power consumption (IoT nodes and network) and traffic mean latency resulting from embedding all the BPs are calculated.

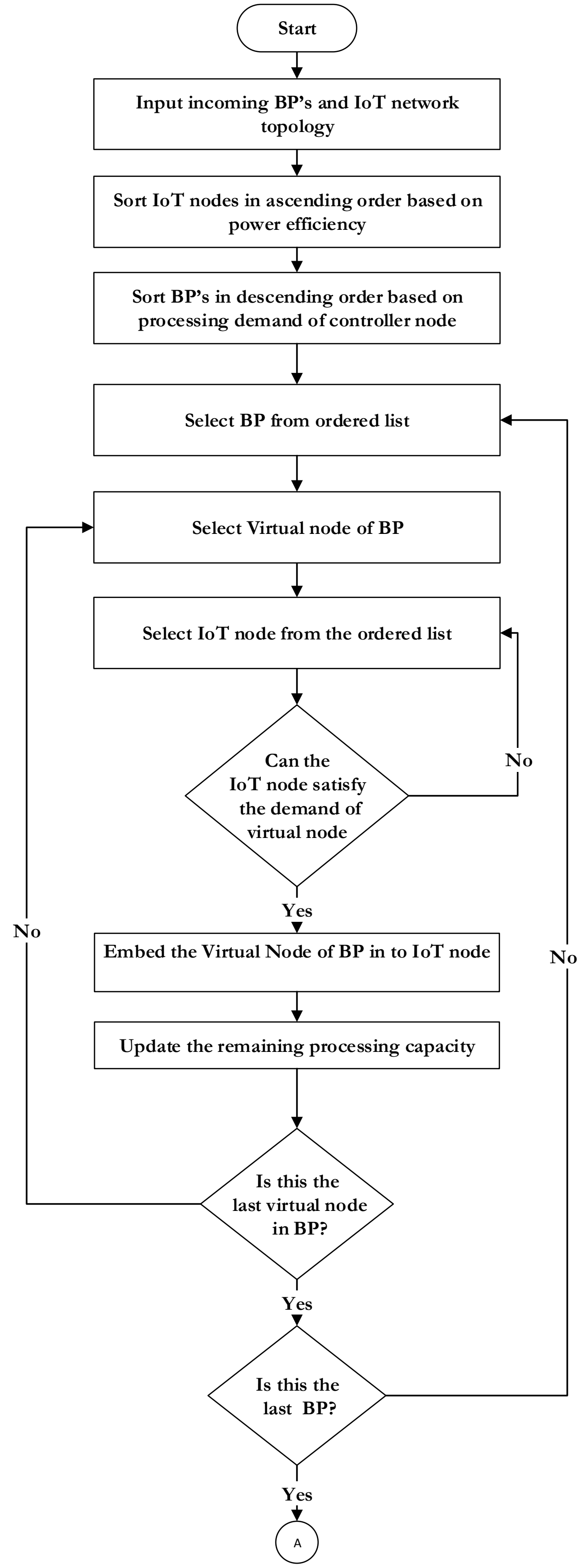

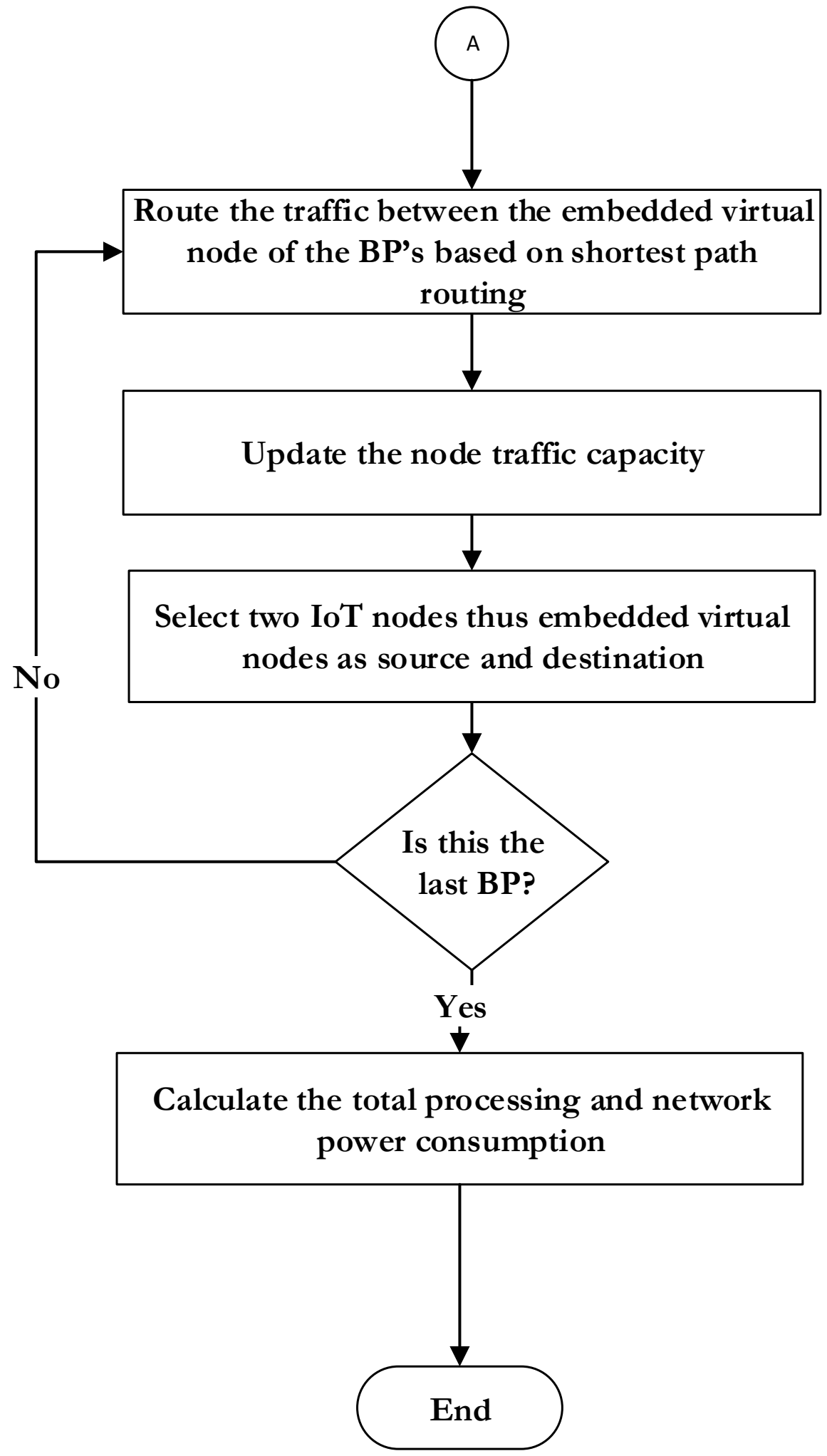

Figure 20: Heuristic Flowchart.

Figure 9 to 12 show that the performance of the RESE heuristic approaches that of the sequential energy efficient MILP model for embedding across different zones. Table 2 summarises the average performance gap between the RESE heuristic and the sequential model.

| | 2 BP's | | 4 BP's | | 6 BP's | |
|---|---|---|---|---|---|---|
| | Sequential MILP | Real time Heuristic | Sequential MILP | Real time Heuristic | Sequential MILP | Real time Heuristic |
| Processing Power | 35 | 23 | 48 | 44 | 96 | 56 |
| Network Power | 420 | 684 | 672 | 736 | 987 | 1149 |

Table 2: Power consumption gap between the RLSE heuristic and the sequential model.

### E. *Real time low latency service embedding heuristic*

The RLSE heuristic reduces the traffic mean latency by setting a threshold on the node transmission capacity utilisation. When routing the traffic between virtual nodes of a BP, the heuristic does not exceed this threshold which grantees distributing the traffic over multiple links. The flowchart of the RLSE heuristic is given in figure 20. The threshold is set to 60% of the maximum node capacity. Different thresholds were examined and this threshold value was identified as the maximum threshold before the latency per node starts increasing fast.

Figure 13 to 16 show that the performance of the RLSE heuristic approaches that of the sequential low latency MILP model for embedding across different zones. Table 3 summarises the average performance gap between the RLSE heuristic and the sequential model.

| | 2 BP's | | 4 BP's | | 6 BP's | |
|---|---|---|---|---|---|---|
| | Sequential MILP | Real time Heuristic | Sequential MILP | Real time Heuristic | Sequential MILP | Real time Heuristic |
| Traffic mean Latency | 40 | 48 | 76 | 81 | 93 | 106 |

Table3: Traffic mean latency gap between the RLSE heuristic and the sequential model.

## [redacted] SUMMARY

This paper has investigated the power consumption and traffic mean latency of service embedding in the IoT network for a smart building setting and has introduced a framework for their minimisation. The services to be embedded are represented by a virtual topology (virtual nodes and links) following a business processes workflow dictated by the SOA paradigm. We developed a MILP framework and a real-time heuristic to optimise the selection of IoT nodes to embed the virtual nodes; and to route the traffic between virtual nodes considering three different objective functions: (i) minimising the total power consumption, (ii) minimising traffic mean latency, (iii) minimising both total power consumption and traffic mean latency in multi-objective manner.

We considered embedding BPs where all the sensor and actuator nodes exist in the same geographical zone and also considered embedding across different zones. We also studied embedding with and without constraints on the coexistence of virtual nodes in the same IoT node.

We used the MILP model to optimise the embedding in two scenarios: (i) re-provisioning scenario where each time a new BPs arrives, previously embedded BPs are re-embedded, (ii) sequential embedding where arriving BPs are embedded without interrupting the existing BPs.

In the energy efficient service embedding scenario, the re-provisioning scenario produces higher average power saving compared with the sequential embedding scenario. In the low latency service embedding scenario, re-provisional embedding reduced the average traffic mean latency compared with the sequential embedding scenario. The multi-objective optimisation shows that it is possible to optimise the embedding of BPs to achieve high optimality of 91% for both power savings and traffic latency.

## [redacted] ACKNOWLEDGEMENTS

The authors would like to acknowledge funding from the Engineering and Physical Sciences Research Council (EPSRC), INTERNET (EP/H040536/1), STAR (EP/K016873/1) and TOWS (EP/S016570/1) projects. Mr. Haider Al-Shammari would like to thank the Higher Committee for Education Development (HCED) for funding his scholarship.

## [redacted]REFERENCES